\documentclass[floats,11pt,floatfix,showpacs,amssymb,amsmath,prd,superscriptaddress,nofootinbib, aps]{revtex4-2}
\usepackage[utf8]{inputenc}
\usepackage{hyperref, amsmath, amsfonts}
\usepackage{graphicx}
\usepackage{aas_macros}
\hypersetup{colorlinks,bookmarksnumbered,
citecolor=cyan, urlcolor=magenta}

\providecommand{\adsurl}[1]{\href{#1}{ADS}}
\usepackage{natbib}
\usepackage[usenames, dvipsnames]{xcolor}
\usepackage{booktabs}
\usepackage{multirow}
\usepackage{amssymb}
\usepackage{empheq}
\usepackage{slashed}
\usepackage{caption}
\usepackage{cellspace}
\usepackage{physics}
\usepackage{subcaption}
\usepackage{mathtools}
\usepackage{ulem}
\usepackage{makecell}
\usepackage{tabularx}
\usepackage{footnote}
\usepackage[export]{adjustbox}

\usepackage{xspace}

\newcommand{\kHz}{\,{\rm kHz}}
\newcommand{\Hz}{\,{\rm Hz}}

\newcommand{\km}{\,{\rm km}}

\newcommand{\yr}{\,{\rm yr}}
\newcommand{\kms}{{\rm km/s}}
\newcommand{\kg}{{\rm kg}}

\newcommand{\DM}{{\rm DM}}
\newcommand{\Doppler}{{\rm Doppler}}
\newcommand{\Shapiro}{{\rm Shapiro}}
\newcommand{\Einstein}{{\rm Einstein}}
\newcommand{\SNR}{{\rm SNR}}
\newcommand{\FSR}{{\rm FSR}}
\newcommand{\peak}{{\rm peak}}
\newcommand{\tidal}{{\rm tidal}}
\renewcommand{\th}{{\rm th}}
\renewcommand{\mid}{{\rm mid}}
\newcommand{\fifth}{{\rm fifth}}

\renewcommand{\vec}[1]{\boldsymbol{\mathbf{#1}}}

\newcommand{\unit}[1]{\vec{\hat{#1}}}

\begin{document}
\hspace{5.2in} \mbox{CALT-TH/2023-021}
\title{Macroscopic Dark Matter Detection with Gravitational Wave Experiments}

\author{Yufeng Du}
\email{yfdu@caltech.edu}
\author{Vincent S. H. Lee}
\email{szehiml@caltech.edu}
\author{Yikun Wang}
\email{yikunw@caltech.edu}
\author{Kathryn M. Zurek}
\email{kzurek@caltech.edu}
\affiliation{Walter Burke Institute for Theoretical Physics, California Institute of Technology, Pasadena, CA 91125, USA}
\date{\today}

\preprint{CALT-TH/2023-021}

\begin{abstract}
We study  signatures of macroscopic dark matter (DM) in current and future gravitational wave (GW) experiments. Transiting DM with a mass of $\sim10^5-10^{15}$ kg that saturates the local DM density can be potentially detectable by GW detectors, depending on the baseline of the detector and the strength of the force mediating the interaction. In the context of laser interferometers, we derive the gauge invariant observable due to a transiting DM, including the Shapiro effect (gravitational time delay accumulated during the photon propagation), and adequately account for the finite photon travel time within an interferometer arm. In particular, we find that the Shapiro effect can be dominant for short-baseline interferometers such as Holometer and GQuEST. We also find that proposed experiments such as Cosmic Explorer and Einstein Telescope can constrain a fifth force between DM and baryons, at the level of strength $\sim 10^3$ times stronger than gravity for, {\it e.g.}, kg mass DM with a fifth-force range of $10^6$ m. 

\end{abstract}

\maketitle
\newpage
\tableofcontents
\newpage

\section{Introduction}
\label{introduction}

Astrophysical and cosmological evidence points to the existence of dark matter (DM), but little has been determined about its microscopic nature, with even its possible mass consistent with observation in the large range of $10^{-22}$ eV to $10^4\,M_{\odot}$ (see Ref.~\cite{green2022snowmass} for a recent review). Direct detection of dark matter in terrestrial experiments has focused on DM particles whose interactions with the Standard Model particles are determined by the DM abundance in the Universe.  Such DM typically has mass $\lesssim 340 \mbox{ TeV}$~\cite{Griest:1989wd}, and has been the subject of a range of experiments searching both for single particle interactions (see Ref.~\cite{mitridate2022snowmass} for a review) and collective wavelike phenomena \cite{2022arXiv220314915A,2023PhRvD.107e5002B}.

On the other hand, the direct detection of ultraheavy dark matter (UHDM) is relatively unexplored, with primordial black holes (PBHs)~\cite{Carr:2016drx} being the most well-studied DM candidate in this category. While unitarity bounds limit DM production through thermal mechanisms above $\sim100$~TeV, UHDM can be a composite state synthesized in a way similar to Standard Model nuclei~\cite{Wise:2014ola,Wise:2014jva,Gresham:2017cvl,Gresham:2017zqi}.  Such UHDM can be searched for by direct scattering~\cite{Coskuner:2018are} or quantum mechanical sensors~\cite{thewindchimecollaboration2022snowmass,Carney:2020xol}. %

In this work, we consider the detection of UHDM beyond $M_\mathrm{Pl}$ via long-range forces, whether gravity or a new fifth force between baryons and DM. Alongside LIGO's success in detecting gravitational waves (GWs) from a binary black hole merger~\cite{LIGO2016}, a myriad of laser interferometer experiments are either in operation or are planned to commence operation in the near future~\cite{Aggarwal2021}. DM transiting in the solar system produces a weak gravitational potential, and can in principle be observed by laser interferometers. These effects have been analyzed previously in the context of pulsar timing arrays (PTAs)~\cite{Siegel:2007fz,Baghram:2011is, Ramani:2020hdo, Lee:2020wfn, Lee_2021, Gresham_2023}. In addition, laser interferometer experiments with shorter ($\sim$ m) baselines, designed to measure quantum gravity signature in causal diamonds (see Ref.~\cite{Zurek:2022xzl} for a review), can also be sensitive to transiting DM. This includes the past experiment Holometer~\cite{Holometer:2015tus, Holometer, Holometer:2017}, an upcoming experiment commissioned by Caltech and Fermilab under the Gravity from the Quantum Entanglement of Space-Time (GQuEST) Collaboration~\cite{McCuller:2022hum}, and a 3D table-top interferometer proposed by the Gravity Exploration Institute at the Cardiff University which probes the transverse correlations of quantum gravity effects \cite{Vermeulen:2020djm}. These GW detectors generally operate at high frequency $\gtrsim 10$ Hz, which corresponds to sensitivity toward DM with mass $M\lesssim$ kg assuming that the DM saturates the local dark matter density.

Detection of UHDM with laser interferometers has been considered elsewhere in the literature~\cite{Hall2016,Jaeckel:2020mqa, Lee:2022tsw}. Other works model GW detectors as simple accelerometers and derive the sensitivity due to transiting DM from mirror acceleration~\cite{Seto:2004zu,Graham_2016, Baum2022,PhysRevD.99.023005}. In this work, we take a more careful approach and formally derive the gauge invariant observable on laser interferometers \cite{Li:2022mvy} from transiting DM.  In addition to the Doppler effect (which is usually the sole effect considered in the literature), the Shapiro delay (gravitational time delay accumulated during the photon propagation) and Einstein delay (gravitational redshift at the detector) are also derived. Moreover, we discuss the statistical formalism for detecting both single events and a stochastic background of events. For other types of GW detectors, we give an overview of those sensitive to transiting DM and project the sensitivity assuming an accelerometer signal. 

Finally, we also consider the possibility that the DM and baryon are coupled with an additional long-range Yukawa interaction, also known as a fifth force~\cite{Adelberger:2003zx}. Such an interaction can arise very generally from an effective Lagrangian with a scalar/vector/tensor mediator between DM and baryons and is only weakly unconstrained by cosmology for heavy DM with force range $\lambda \lesssim 10^6$~m, even with stronger-than-gravity coupling. The existence of a long-range fifth force can have profound implications in DM searches, such as the creation of DM evaporation barriers in celestial bodies~\cite{Acevedo:2023owd}. Various experiments searching for weak equivalence principle violating forces have put constraints on specific models, such as coupling through massive scalars \cite{2022CQGra..39t4010B,2022PhRvD.106i5031B}. Here we consider a more general scenario without the assumption of the specific microscopic interaction. We find that high-frequency detectors are able to constrain the Yukawa coupling constant to be $\lesssim 10^3$ for a force range $\lambda>10^6$ m and $M\sim$ kg within one year of integration time, which is roughly consistent with the findings of Refs.~\cite{Baum2022, Lee:2022tsw}.  %

Our paper is organized as follows. 
In~Section~\ref{sec:signal_description}, we provide a description of the gauge invariant strain from transiting macroscopic DM and discuss various aspects of the signal. In~Section~\ref{sec:signal_derivation}, we provide a detailed derivation of the signal spectrum. In~Section~\ref{sec:stochastic}, we derive the constraints from a stochastic signal, where individual DM might be insufficient to produce a detector signature, but the collective behavior from all DM passing by the detector can be detectable. %
In Section~\ref{sec:sensitivity} we discuss the various experiments we consider in this work and their sensitivity curves to the signal. 
In Section~\ref{sec:discussion} we place our results into context and conclude.

\section{Description of Macroscopic Dark Matter Signals in Interferometry-Based Gravitational Wave Detectors}
\label{sec:signal_description}

Transiting DM induces a gravitational field as a metric perturbation $h_{\mu\nu}$. This produces a strain in a GW detector, $h(t)=\Delta L/L$ (where $L$ is the interferometer arm length), which is the sum of three individual contributions %
\begin{enumerate}
	\item Doppler effect: acceleration of the mirrors 
	\item Shapiro delay: change in the photon travel time within the interferometer arm
	\item Einstein delay: time dilation of the clock proper time (also known as gravitational redshift)
\end{enumerate}
The total strain can be written in the general form
\begin{equation}\label{eqn:total_strain}
	h(t, \unit{n}) = h_{\Doppler}(t, \unit{n}) + h_{\Shapiro}(t, \unit{n}) + h_{\Einstein}(t) \, ,
\end{equation} 
where $\unit{n}$ is the unit vector along the interferometer arm. Note that the Einstein delay does not depend on the arm orientation. A Michelson-Morley laser interferometer consists of two arms and measures the \textit{difference} between the two arms
\begin{equation}\label{eqn:two_arm_strain}
	h(t, \unit{n}_1, \unit{n}_2) = h(t, \unit{n}_1) - h(t, \unit{n}_2) \, ,
\end{equation} 
where $\unit{n}_1$ and $\unit{n}_2$ are the orientation of the two arms respectively. We quickly see that the Einstein delay contribution vanishes for laser interferometers, but can be present in single-arm interferometers, such as PTAs and long-baseline atom interferometers. In the following sections, we will suppress the unit vector dependence for simplicity. We emphasize that individual contributions from Eq.~\eqref{eqn:total_strain} are frame dependent, and only the sum is gauge invariant and is hence an acceptable experimental observable~\cite{Li:2022mvy}. In~Sec.~\ref{sec:signal_derivation}, we will review this gauge invariant time delay and derive the resulting signal in full in the context of a transiting DM.
\begin{figure*}[h]
	\includegraphics[scale=0.75]{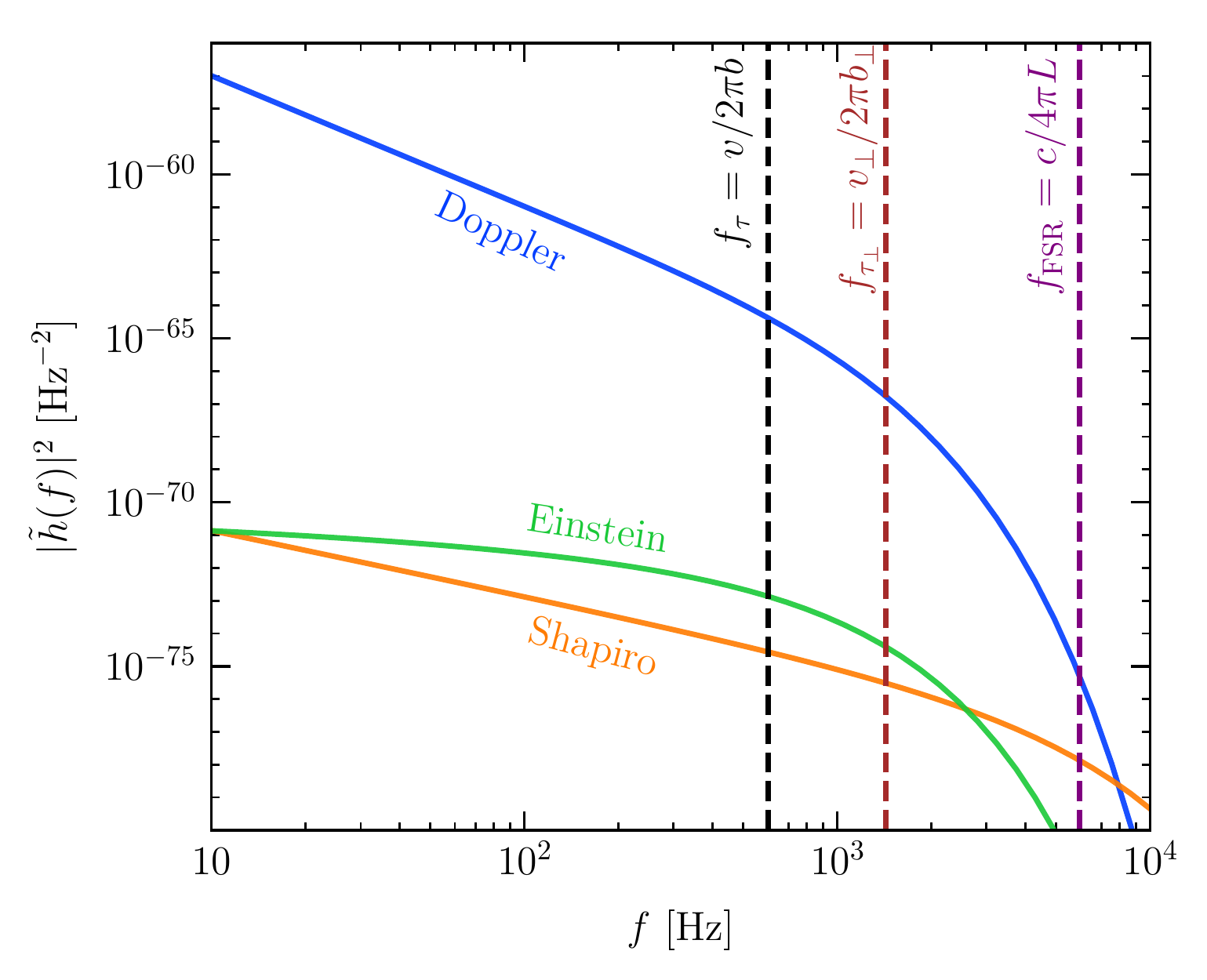} 
	\caption{Signal spectrum $|h_X(f)|^2$ for Doppler effect, Shapiro delay and Einstein delay. Here we choose $M=10^{-4}$ kg, $b=90$ m, $v=340$ km/s, $b_{\perp}=30$ m, $v_{\perp}=270$ km/s, $L=4$ km, and assume a single-arm GW detector for illustration purposes. See the discussion of different length scales in Sec.~\ref{subsec:distance}. The analytic expressions for the signal spectrum are taken from Eq.~\eqref{eqn:doppler_Fourier_psd} (Doppler), Eq.~\eqref{eqn:Shapiro_fourier_magnitude} (Shapiro), and Eq.~\eqref{eqn:Einstein_Fourier_magnitude} (Einstein).}\label{fig:doppler_vs_shapiro_vs_einstein}
\end{figure*}

In Fig.~\ref{fig:doppler_vs_shapiro_vs_einstein} we show the Fourier transformed strain, $\tilde{h}_X(f)\equiv \int dt\exp(-2\pi ift)h_X(t)$, for each contribution ($X$ = Doppler, Shapiro or Einstein) and some choices of the DM mass $M$, velocity $v$, perpendicular velocity $v_\perp$, impact parameter $b$, and perpendicular impact parameter $b_\perp$, as specified in the figure caption. Here the perpendicular components (denoted by the subscript $\perp$) are defined relative to the interferometer arm. Analytic formulas for the spectrum are derived in Sec.~\ref{sec:signal_derivation}. We observe that the signal is a simple power law in the frequency $f$ when $f\lesssim f_{\tau}$, and rapidly drops to zero as $f\gtrsim f_{\tau}$,where $f_{\tau}\equiv 1/( 2\pi \tau)$ is the characteristic frequency of a transiting DM as given by the characteristic timescale $\tau\equiv b/v$, or in the $b\lesssim L/2$ limit of the Shapiro delay, $\tau_{\perp}\equiv b_{\perp}/v_{\perp}$  (see~Sec.~\ref{subsec:distance} for a detailed discussion). In general, the signal spectrum can be parametrized as
\begin{equation}\label{eqn:spectrum_parameterize}
	|\tilde{h}_X(f)|^2 = A_X^2 \left|\tilde{q}_X\left(\frac{f}{f_{\FSR}}\right)\right|^2\left|\tilde{s}_X\left(\frac{f}{f_{\tau}}\right)\right|^2 \, ,
\end{equation} 
where $f_{\FSR}\equiv c/4\pi L$ is the detector's free-spectral-range (FSR) frequency, characterizing the time needed for the photon to complete a roundtrip within the interferometer, $\tilde{q}_X(x)$ is its associated spectral shape, and $\tilde{s}_X(x)$ is the spectral shape of the DM signal. The constant coefficient $A_X$ characterizes the amplitude of the signal. Explicit forms of each component are derived in~Sec.~\ref{sec:signal_derivation}. See~Table~\ref{table:spectrum} for a summary of the analytic expressions.

Most laser interferometers designed to measure GWs such as LIGO utilize Fabry-P\'erot (FP) cavities to increase interaction time between photons and the GW. A direct consequence is that the detector peak sensitivity is displaced from the FSR frequency by the cavity quality factor $Q\gg 1$, {\it i.e.} $f_{\peak}/f_{\FSR}\sim 1/Q$. As we will derive in~Sec.~\ref{sec:signal_derivation}, the effect of finite photon travel time produces corrections to the signal spectrum in powers of $(f/f_{\FSR})$ (Eq.~\eqref{eqn:doppler_Fourier_psd}, Eq.~\eqref{eqn:Shapiro_fourier_magnitude}, and Eq.~\eqref{eqn:Einstein_Fourier_magnitude}), and hence can be safely ignored for these experiments. However, experimental apparatus designed to measure quantum gravity effects such as Holometer and GQuEST generally do not have FP cavities, since the peak frequency of quantum gravity signatures is naturally the frequency associated with the photon travel time in the physical interferometer arm, {\it i.e.} the FSR frequency~\cite{Li:2022mvy}. Hence these quantum gravity detectors generally are most sensitive to signals that peak at $f_{\FSR}$, and the photon travel time within the apparatus cannot be neglected. 

We discuss various important distance scales in Sec.~\Ref{subsec:distance} and suppression effects in Sec.~\Ref{subsec:tidal}. In Sec.~\Ref{subsec:projected_sensitivity} and Sec.~\Ref{subsec:fifth_force} we present and discuss the projected constraints on DM interacting with the detectors gravitationally and with a long-range fifth force, respectively.

\subsection{Distance Scales}
\label{subsec:distance}

\begin{figure*}[h]
	\includegraphics[scale=0.75]{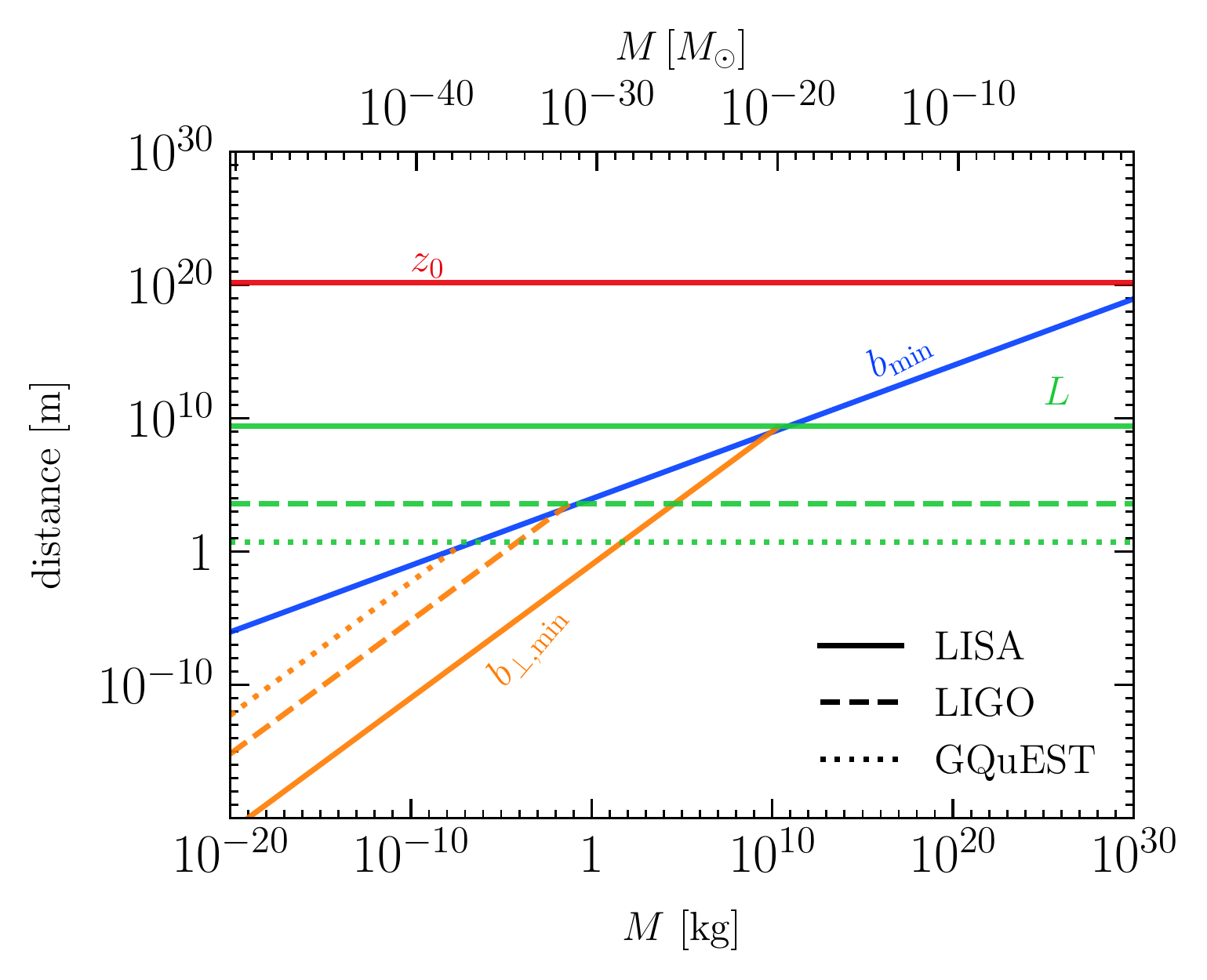} 
	\caption{Relevant distance scales for measuring transiting DM signals as a function of DM mass $M$, assuming $f_{\DM}=1$ and a local DM mass density of $\rho_{\DM}=0.46\,\text{GeV}/c^2/\text{cm}^3$. The length scale for Doppler and Einstein effect is $b_{\min}$, while the length scale for Shapiro is $b_{\min}$ if $b_{\min}\gtrsim L/2$ and $b_{\perp, \min}$ otherwise. Note that $b_{\perp, \min}$ depends on the length scale of the detector baseline, for which we choose three experiments with diverse baselines (LISA, LIGO, and GQuEST) for illustration purposes. For reference we also show a typical pulsar-Earth distance, $z_0 \sim 5$ kpc, which is the largest distance scale for PTA searches when $M<10^2\,M_{\odot}$.}\label{fig:distances}
\end{figure*}

The signal timescale and the corresponding frequency of a transiting DM depend on its distance of closest encounter with the detector. The distribution of DM around the detector gives rise to a relation between the distance scale and the DM density. Here we summarize various relevant distance scales for the DM signals. 
For Doppler and Einstein delay, since the DM effect only acts on a point (the mirrors or the clock), the relevant distance scale is $b$. For Shapiro delay, if the DM is sufficiently distant ($\gtrsim L/2$) from the detector, then the entire interferometer arm is effectively a point and the relevant distance is still $b$. However, for nearby DM ($b\lesssim L/2$), the relevant scale for Shapiro delay is the DM's closest encounter to any point along the arm rather than a specific point, denoted as $b_{\perp}$ (note that $b_{\perp}\leq b$ by definition). The local statistical distribution of $b$ and $b_{\perp}$ of DM has been studied and derived in the appendix of Ref.~\cite{Dror:2019twh}. In particular, the $90^{\mathrm{th}}$ percentile minimum DM impact parameters, $b_{\min}$ and $b_{\perp,\min}$, are given by 
\begin{align}\label{eqn:b_min}
	b_{\min} &= \sqrt{-\frac{\log(1-p)}{\pi n\bar{v}T}}= 9\,\mathrm{km}\left(\frac{M}{\kg}\right)^{1/2}f_{\DM}^{-1/2}\left(\frac{340\,\kms}{\bar{v}}\right)^{1/2}\left(\frac{\yr}{T}\right)^{1/2},  \nonumber \\
	b_{\perp,\min} &= -\frac{\log(1-p)}{n\bar{v}_{\perp}TL} = 300\,\mathrm{km}\left(\frac{M}{\kg}\right)f_{\DM}^{-1}\left(\frac{\km}{L}\right)\left(\frac{270\,\kms}{\bar{v}_{\perp}}\right)\left(\frac{\yr}{T}\right)\, ,
\end{align} 
where $p$ is the percentile of the minimal impact parameters (taken to be $0.9$ for the numerical estimate above), $n=\rho_{\DM}f_{\DM}/M$ is the local DM number density, with $\rho_{\DM}=0.46\,\text{GeV}/c^2/\text{cm}^3$, $f_{\DM}$ the DM fraction in mass $M$, $\bar{v}$ and $\bar{v}_{\perp}$ are the average DM velocity and perpendicular velocity respectively, and the estimate for $b_{\perp,\min}$ only holds when $b_{\min}<L$. In Fig.~\ref{fig:distances} we show the distance scales from LISA, LIGO, and GQuEST for different choices of DM mass $M$. As will be discussed in the next subsection and carefully verified in~Sec.~\ref{sec:signal_derivation}, only DM with impact parameter $b\lesssim L$ can potentially generate sizable signals. We see that $b_{\min}$ and $b_{\perp,\min}$ coincide at $\sim L$. At those lower mass ranges, the Shapiro effect is boosted by the fact that $b_{\perp}<b$, but as we will see in the next section (cf. Eq.~\eqref{eqn:Shapiro_fourier_magnitude}), the Shapiro delay in the $b<L/2$ limit suffers a suppression factor of $v/c$. These competing factors lead to the dominance of the Doppler effect in most experiments, but the Shapiro effect has a slight edge for specific values of $L$ and $M$.

For reference, we also show the typical distance between Earth and a pulsar observed in PTA experiments, $z_0\sim 5$ kpc. For $M\lesssim 10\,M_{\odot}$, which is the mass range considered in most previous works on PTA~\cite{Dror:2019twh, Ramani:2020hdo, Lee:2020wfn, Lee_2021, Gresham_2023}, $z_0$ is the largest distance scale. This shows a natural mass cutoff when extending the previous PTA results for DM with mass $M\gtrsim 10 M_{\odot}$, as we do not expect DM to give measurable signatures in PTAs if $b\gtrsim z_0$.

\subsection{Tidal and $Q$-Suppression}
\label{subsec:tidal}

Laser interferometers are mostly sensitive to DM with impact parameter $b\lesssim L$. The main reason is that if $b\gtrsim L$, the peak frequency of the signal $f_{\tau}\lesssim 2(v/c)f_{\FSR}\sim 10^{-3}f_{\FSR}$. Thus unless the $Q$ factor is sufficiently large ($Q>10^3$), the peak DM frequency typically falls below the sensitivity window of the GW detector. Notice that the DM signal drops exponentially above such a signal peak frequency. In addition, the strain due to DM with $b\gtrsim L$ suffers suppression from two other effects, which we will see explicitly from the derivation in Sec.~\ref{sec:signal_derivation}.:
\begin{itemize}
	\item {\em Tidal suppression.} Since an interferometer measures differential quantities, when $b\gtrsim L$, the interferometer behaves like a dipole under a gravitational field, and thus the signal can be suppressed by a factor of $L/b$, which is commonly known as the tidal effect. 
	\item {\em $Q$-suppression.} When $b\gtrsim L$, the signal can evade tidal suppression since the interferometer measures the differential strain at slightly different times, creating an envelope in Fourier space that is peaked at $f_{\FSR}$~\footnote{Effects of finite photon travel time are discussed in the context of gravitational waves in Refs.~\cite{Larson:1999, Cornish:2002, Rakhmanov:2004, Rakhmanov:2008} and ultralight DM in Refs.~\cite{Morisaki:2018, Morisaki2021}} . When the signal is evaluated at the detector peak sensitivity $f_{\rm peak}$, the signal picks up a factor of $1/Q$.
\end{itemize} 
When the signal with $b>L$ is evaluated at frequency $f=f_{\tau}$, since $(1/Q)(f_{\tau}/f_{\rm peak})=f_{\tau}/f_{\FSR}=2(v/c)(L/b)$, we see that the $Q$-suppressed term is always weaker than the tidally suppressed term for laser interferometers as $v\ll c$. However, for some types of interferometers, such as atom interferometers, the speed of the probe can be much slower than the DM speed. In that case $f_{\tau}>f_{\FSR}$ is possible even when $b>L$, and the $Q$-suppressed term can dominate over the tidally suppressed term. We leave the detailed treatment of these types of experiments for future work. In Table~\ref{table:suppression} we summarize the suppression factors for $b\gtrsim L$ for all contributions in Eq.~\eqref{eqn:total_strain}, and for both one-arm and two-arm interferometers, which will be justified in later sections.

\subsection{Projected Sensitivity}
\label{subsec:projected_sensitivity}

To set the projected sensitivity for various current and future GW detectors, we assume that the detector noise is stationary and Gaussian with a one-sided power spectral density $S_n(f)$ (in units of Hz$^{-1}$). The deterministic signal-to-noise ratio (SNR), assuming optimal filtering in a matched filtering procedure, is given by~\cite{Maggiore:2007ulw}
\begin{equation}\label{eqn:deterministic_SNR}
	\SNR_{\mathrm{det}}^2 = 4\int_0^{\infty}df\frac{|\tilde{h}(f)|^2}{S_n(f)} \, .
\end{equation} 
For narrowband GW detectors, the deterministic SNR can be approximated as $\SNR_{\mathrm{det}}^2 \approx (4\Delta f/S_n)|\tilde{h}(f_{\peak})|^2$, where $\Delta f$ is the narrow frequency bandwidth.

On the signal side, the spectrum in Eq.~\eqref{eqn:spectrum_parameterize} can be greatly simplified assuming $b<L$ for the purpose of computing the SNR, which takes much simpler forms truncated at $f=f_{\tau}$. Here we show the approximated form of the signal strain by quoting the results from Sec.~\ref{sec:signal_derivation} and taking the $f\ll f_{\tau}$ limit of Eq.~\eqref{eqn:doppler_Fourier_psd}, Eq.~\eqref{eqn:Shapiro_fourier_magnitude} and Eq.~\eqref{eqn:Einstein_Fourier_magnitude} 
\begin{align}\label{eqn:simplified_spectrum}
	|\tilde{h}_{\Doppler}(f)|^2 &\approx \frac{4}{3}\left(\frac{8GML}{c^2bv}\right)^2\left(\frac{f_{\FSR}}{f}\right)^4\cos^2\left(\frac{f}{2f_{\FSR}}\right)\Theta(f_{\tau}-f), \nonumber \\
	|\tilde{h}_{\Shapiro}(f)|^2 &\approx \left(\frac{8\pi GM}{c^3}\right)^2\left(\frac{f_{\FSR}}{f}\right)^2\cos^2\left(\frac{f}{4f_{\FSR}}\right)\Theta(f_{\tau_{\perp}}-f), \nonumber \\
    |\tilde{h}_{\Einstein}(f)|^2 &\approx \left(\frac{8GM}{c^2v}\right)^2\left(\frac{f_{\FSR}}{f}\right)^2\sin^2\left(\frac{f}{2f_{\FSR}}\right)\log^2\left(\frac{f}{f_{\tau}}\right)\Theta(f_{\tau}-f)  \, ,
\end{align} 
for one-arm detectors. For two-arm interferometers $|\tilde{h}_{\Doppler}(f)|^2$ should pick up a factor of $4\sin^2(\Delta\theta/2)$, where $\Delta \theta$ is the angle between the two arms, $|\tilde{h}_{\Shapiro}(f)|^2$ does not receive a correction when $b<L/2$ and $\Delta\theta\sim\mathcal{O}(1)$, and $|\tilde{h}_{\Einstein}(f)|^2=0$. The simplified spectrum is very accurate for the lower mass range where $b<L$, but can underestimate the upper limits on $f_{\DM}$ by $\lesssim$ 4 orders of magnitude on the higher mass range. 

The 90$^{\text{th}}$ percentile upper limit on $f_{\DM}$ is derived by requiring $b_{\rm min}>L$ (in the high mass limit), and the 10$^{\text{th}}$ percentile SNR to be less than 2 (in the low mass limit), where the SNR is produced by DM with impact parameter given by~Eq.~\eqref{eqn:b_min}. Computing $\SNR_{\det}$ in Eq.~\eqref{eqn:deterministic_SNR} using the simplified signal strain in~Eq.~\eqref{eqn:simplified_spectrum}, the constraints on $f_{\DM}$ are then roughly given by
\begin{align}\label{eqn:f_constraint_simple}
	&f^L_{\DM,\Doppler} 
	\lesssim 2\times 10^{16} \left(\frac{\kg}{M}\right)\left(\frac{\yr}{T}\right)\left(\frac{\bar{v}}{340\,\kms}\right)\left(\frac{\km}{L}\right)^2\left(\frac{1}{Q}\right)^4\left(\frac{S_n}{10^{-46}\,\Hz^{-1}}\right)\left(\frac{\kHz}{\Delta f}\right),\nonumber \\
& f^R_{\DM,\Doppler} 
 \lesssim 80 \left(\frac{M}{\kg}\right)\left(\frac{\yr}{T}\right)\left(\frac{340\,\kms}{\bar{v}}\right)\left(\frac{\km}{L}\right)^2, \nonumber \\
& M^L_{\Shapiro} 
\lesssim 5\times 10^{9}\,\kg \left(\frac{1}{Q}\right)\left(\frac{S_n}{10^{-46}\,\Hz^{-1}}\right)^{1/2}\left(\frac{\kHz}{\Delta f}\right)^{1/2}, \nonumber \\
&f^R_{\DM,\Shapiro} \lesssim 3\times 10^2 \left(\frac{M}{\kg}\right)\left(\frac{\yr}{T}\right)\left(\frac{270\,\kms}{\bar{v}_{\perp}}\right)\left(\frac{\km}{L}\right)^2 \, .
\end{align} 
Here the superscripts ``$L$" and ``$R$" denote the low and high mass regions of the parameter space, respectively. Note that in the low mass regime, the SNR for the Shapiro effect becomes independent of $f_{\DM}$, for which case we show the constraint on the DM mass $M$ instead. 

\begin{figure*}[t]
	\includegraphics[scale=0.7,left]{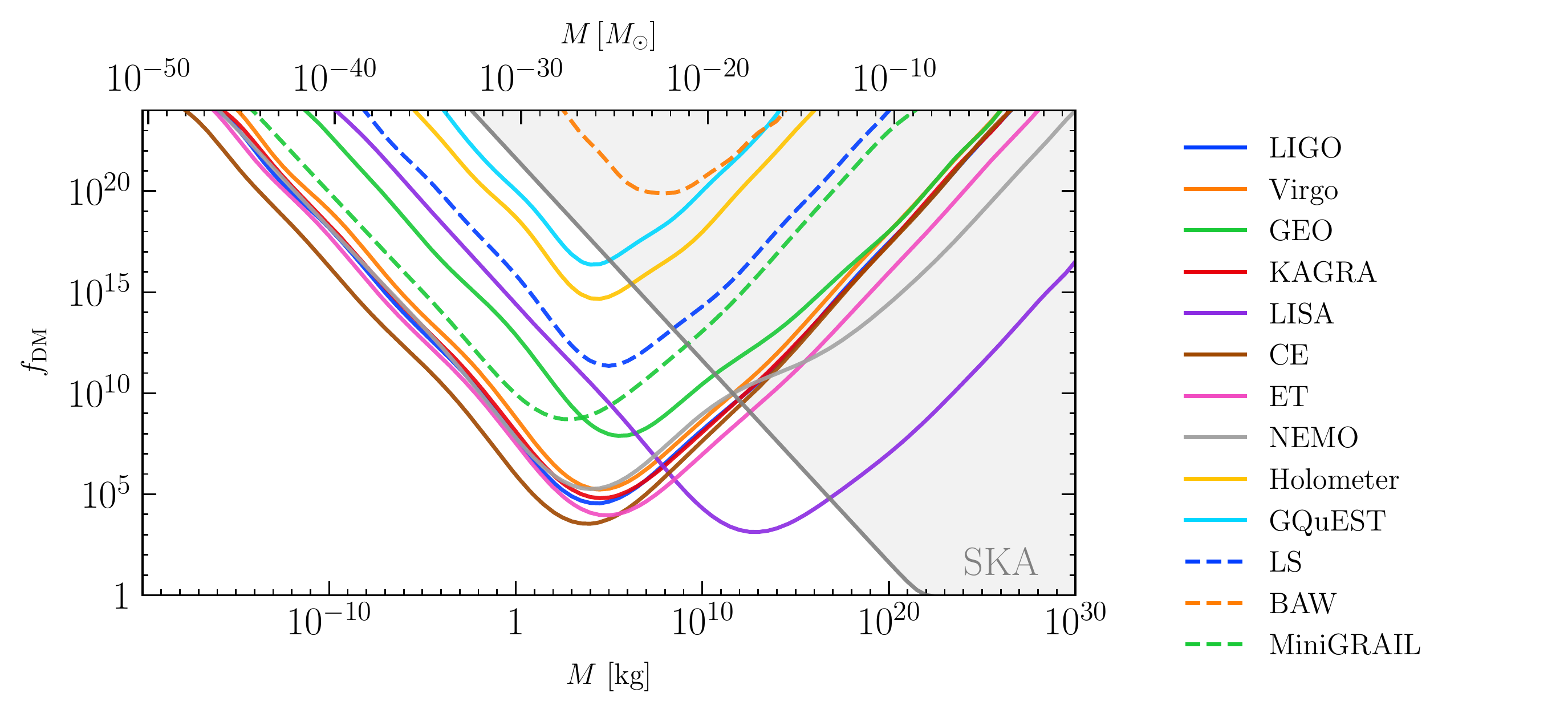} 
	\caption{Projected 90$^{\th}$-percentile upper limits on transiting DM fraction from several existing and proposed GW detectors based on laser interferometry, assuming $T=1$ yr of observation time and local DM density $\rho_{\DM}=0.46$ GeV/$c^2/$cm$^3$. The limits are derived by setting the $10^{\th}$-percentile SNR defined in Eq.~\eqref{eqn:deterministic_SNR} to be two, and the DM initial conditions are sampled using a Monte Carlo simulation. Projections from other types of high-frequency GW experiments are shown with dashed colored lines. See Sec.~\ref{sec:sensitivity} for a description of the experiments.}\label{fig:reach}
\end{figure*}

In Fig.~\ref{fig:reach} we show the projected constraints for several existing and proposed GW experiments based on laser interferometry, assuming $T=1$ yr of observation time. These experiments are discussed in~Sec.~\ref{sec:sensitivity} in more detail, with the noise spectral densities plotted together in~Fig.~\ref{fig:Sh}. We derive the projections using a Monte Carlo simulation to sample the DM initial conditions, compute the SNR with the exact strain as derived and shown in~Sec.~\ref{sec:signal_derivation}, and set the SNR to 2. In the Monte Carlo simulation, the DM impact parameters are randomly sampled and properly normalized to the local DM density, while the velocity distribution is taken to be the Standard Halo Model (SHM), {\it i.e.} an isotropic Maxwell-Boltzmann distribution with $v_0=220$ km/s, boosted by the solar system speed $v_0=220$ km/s, and truncated at the escape velocity $v_{\mathrm{esc}}=600$ km/s. The mean DM velocities are $\bar{v}=340$ km/s and $\bar{v}_{\perp}=270$ km/s~\cite{Ramani:2020hdo}. The trajectories of DM, dependent on time, are then parametrized by the impact parameters, velocities, and arrival time of the DM (see Eq.~\eqref{eqn:DM_trajectory_b} and Eq.~\eqref{eqn:DM_trajectory_b_perp} for explicit forms and discussion in Sec.~\ref{sec:signal_derivation}). The constraints are reported in terms of $f_{\DM}$, defined to be the fraction of DM as transiting point masses. We find that for laser interferometers, the Doppler effect is dominant over the Shapiro delay except for the high mass range in Holometer and GQuEST (which appears as bumps in the constraint curves). We see that gravitational signals from transiting DM are out of reach for laser interferometry-based GW detectors, even if the DM local density is saturated. However, if there exists an additional long-range fifth force between DM and baryonic matter, GW detectors can be sensitive to a fifth force $\sim 10^3$ times stronger than gravity within a year of observation, which will be elaborated in Sec.~\ref{subsec:fifth_force}. On the same plot, we show projections from other types of high-frequency GW detectors, which are modeled as accelerometers for simplicity. More specifics are explained in Sec.~\ref{sec:sensitivity}.

\begin{table}[!htb]
    \begin{minipage}{.5\linewidth}
      \centering
        \begin{tabular}{|c||c|c|}
            \hline
            \multicolumn{3}{|c|}{$|\tilde{h}_X(f)|^2 = A_X^2 \left|\tilde{q}_X\left(\frac{f}{f_{\FSR}}\right)\right|^2\left|\tilde{s}_X\left(\frac{f}{f_{\tau}}\right)\right|^2$}                                                 \\ \hline \hline
                                           & one-arm               & two-arm               \\ \hline \hline
            Doppler                        & Eq.~\eqref{eqn:doppler_Fourier_psd}           &  $4\sin^2(\Delta\theta/2)\times$ Eq.~\eqref{eqn:doppler_Fourier_psd}                 \\ \hline
            Shapiro                        & Eq.~\eqref{eqn:Shapiro_fourier_magnitude}     &  Eq.~\eqref{eqn:Shapiro_fourier_two_arm}                                             \\ \hline
            Einstein                       & Eq.~\eqref{eqn:Einstein_Fourier_magnitude}    &  $0$                       \\ \hline
        \end{tabular} \caption{Reference equations for Doppler, Shapiro, and Einstein signal spectrum. A skeleton form of the spectrum is given by Eq.~\eqref{eqn:spectrum_parameterize}. Here $\Delta\theta$ is the angular separation between the two interferometer arms.}\label{table:spectrum}
    \end{minipage}%
    \begin{minipage}{.5\linewidth}
      \centering
        \begin{tabular}{ |c||c|c| }
			\hline
            \multicolumn{3}{|c|}{suppression factor when $b>L$}                                                 \\ \hline \hline
			& one-arm & two-arm \\ \hline \hline
			Doppler & tidal + $Q$ & tidal + $Q$ \\ \hline
			Shapiro & 1 & tidal + $Q$ \\ \hline
			Einstein & 1 & $\to 0$ \\
			\hline
		\end{tabular}\caption{Suppression factors for different contributions to the strain when $b\gtrsim L$. Here the notation ``tidal + $Q$" denotes that the signal is the sum of two terms, suppressed by the tidal and $Q$ factor respectively.}\label{table:suppression}
    \end{minipage} 
\end{table}

\subsection{Fifth Force}
\label{subsec:fifth_force}
\begin{figure*}[h]
	\includegraphics[scale=0.55]{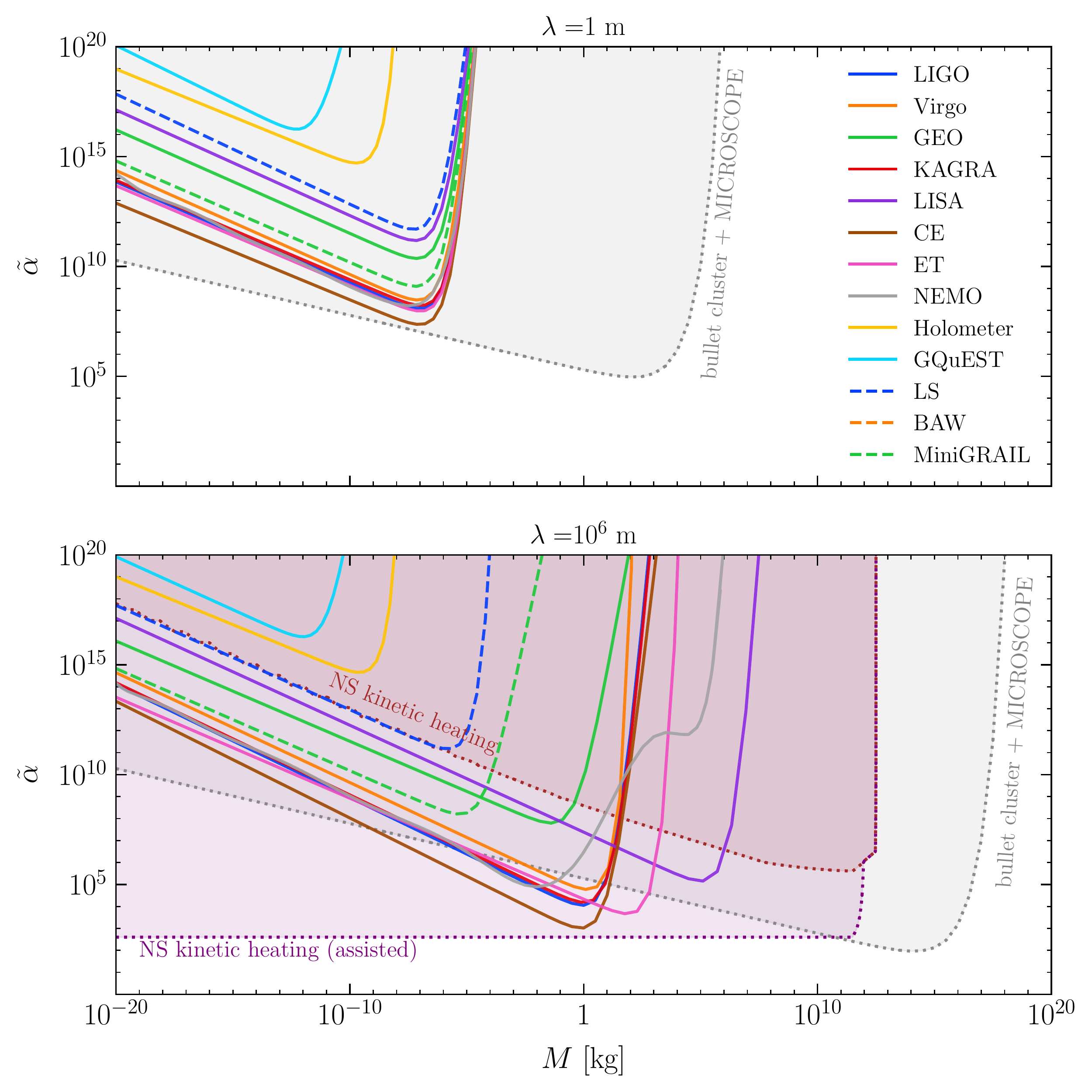} 
	\caption{Projected 90$^{\th}$-percentile upper limits on the fifth-force Yukawa parameter from several existing and proposed GW detectors based on laser interferometry, assuming $T=1$~yr of observation time and two choices of force range, $\lambda=1$ m and $\lambda=10^6$ m. Existing constraints are shown in dotted lines. The gray line is the combined constraint from bullet cluster observation~\cite{Spergel_2000, Kahlhoefer_2013} and the MICROSCOPE experiment~\cite{Berg__2018, Fayet_2019}, while the purple (red) lines are constraints from neutron star kinetic heating~\cite{Gresham_2023} with (without) additional short-range DM-baryon interactions to facilitate energy transfer. The limits are derived by setting the $10^{\th}$-percentile SNR defined in Eq.~\eqref{eqn:deterministic_SNR} to be 2, and the DM initial conditions are sampled using a Monte Carlo simulation. Projections from other types of high-frequency GW experiments are shown with dashed colored lines. See Sec.~\ref{subsec:fifth_force} for a summary of existing fifth-force constraints, and Sec.~\ref{sec:sensitivity} for a description of the experiments.}\label{fig:fifth_force_reach}
\end{figure*}

In the presence of a long-range DM-baryon Yukawa force (also known as a fifth force), the potential can be written as
\begin{equation}\label{eqn:fifth_force}
	\Phi_{\fifth}(r) = -\tilde{\alpha}\frac{GM}{r}e^{-r/\lambda} \, ,
\end{equation} 
where $\tilde{\alpha}$ is the coupling strength (normalized against gravity), and $\lambda$ is the force range. The effect of the fifth force can be estimated using the same signal spectrum in Eq.~\eqref{eqn:spectrum_parameterize} but truncated at $b\gtrsim \lambda$ for Doppler effect (note that a fifth force coupled to $B$ or $B-L$ induces no Shapiro delay). In Fig.~\ref{fig:fifth_force_reach} we show the resulting projected constraints from the Monte Carlo simulation on $\tilde{\alpha}$ for $\lambda = 1$ m and $\lambda = 10^6$ m, alongside several existing fifth-force constraints. The finite force range introduces a cutoff mass corresponding to $b_{\min}\sim \lambda$. We observe that constraints on $\tilde{\alpha}$ for experiments with long baselines such as LISA and LIGO significantly weaken when the force range $\lambda$ drops below the interferometer length. However, experiments with shorter baselines such as Holometer and GQuEST are less insensitive to the shorter force range as long as $\lambda \gtrsim 1$ m, since the peak sensitivity of these experiments corresponds to the $b\sim 1$ m scale.

Astrophysical constraints on the DM-baryon fifth force include weak equivalence principle (WEP) tests, which measure the differential acceleration of two baryonic bodies toward the Galactic Center. Several existing WEP analyses include perihelion precession (Sun-Mercury)~\cite{Sun_2019}, binary pulsar (NS-white dwarf)~\cite{Shao_2018}, lunar laser ranging (Earth-Moon) and torsion pendula (Be-Ti, Be-Al)~\cite{Wagner_2012}, but have been shown to be subdominant (upper limits on $\tilde{\alpha}>10^{20}$) for $\lambda<10^6$ m~\cite{Gresham_2023}. Observation of neutron star (NS) surface temperature can also place upper limits on the Yukawa coupling constant since a large DM-baryon interaction leads to a high NS temperature due to kinetic heating~\cite{Gresham_2023}. An indirect bound on DM-baryon interaction comes from combining~\cite{Coskuner:2018are} upper limits on DM self-interaction from bullet cluster observation~\cite{Spergel_2000, Kahlhoefer_2013}, and bounds on baryon-baryon fifth force measured in Eötvös experiments such as MICROSCOPE~\cite{Berg__2018, Fayet_2019}, which is shown in Fig.~\ref{fig:fifth_force_reach}. While the bullet cluster + MICROSCOPE bound is dominant over the GW detector bounds for most mass ranges, if only a subcomponent (say 1\%) of DM is charged under the fifth force, then the bullet cluster bound on DM self-interaction does not exist, while the GW detector bounds will only deteriorate linearly with the subcomponent fraction.

\section{Derivation of the Signal}
\label{sec:signal_derivation}

The observable in a laser interferometer experiment is the total photon roundtrip time within an interferometer arm. In Ref.~\cite{Li:2022mvy}, this quantity is written as a sum of three separate components, corresponding to the Doppler effect, the Shapiro effect, and the Einstein time delay. The sum has been explicitly shown in Ref.~\cite{Li:2022mvy} to be invariant under general gauge transformations. Here we follow Ref.~\cite{Li:2022mvy} and briefly summarize this gauge invariant quantity. Then, we restrict ourselves to the scenario of a pointlike transiting DM and explicitly derive the strain for each effect. 

In the presence of a general metric perturbation
 \begin{equation}
 	ds^2= -(1-\mathcal{H}_0)dt^2 + (1+\mathcal{H}_2)dr^2+2\mathcal{H}_1dtdr..., 
 \end{equation}
the total photon travel time within a roundtrip, $T_{\gamma}$, in an interferometer centering at the origin can be computed by including effects from the clock rate change, mirror motion, and gravitational redshift in the photon geodesic \cite{Li:2022mvy}:
\begin{equation}
\begin{split}
    T_{\gamma}(t)=T_{\gamma}^\text{\rm out}(t) + T_{\gamma}^\text{\rm in}(t) =\delta \tau &+\frac{1}{c}\int_{r_\text{\rm M}\left(t,0\right)}^{L+r_\text{\rm M}(t+\frac{L}{c},L)} dr \left(1+\frac{1}{2}\mathcal{H}^\text{\rm out}\left(t+\frac{r}{c},r\right)\right) \\
    &- \frac{1}{c}\int_{L+r_\text{\rm M}\left(t+\frac{L}{c},L\right)}^{r_\text{\rm M}\left(t+\frac{2L}{c},0\right)} dr \left(1+\frac{1}{2}\mathcal{H}^\text{\rm in}\left(t+\frac{2L-r}{c},r\right)\right)
\end{split}
\end{equation}
where $t$ is the time when the photon leaves from the beamsplitter. Here $\mathcal{H}^\text{\rm out/in}\equiv \mathcal{H}_0+\mathcal{H}_2\pm 2\mathcal{H}_1$, $r_M(t',r')$ denotes the mirror position at time $t'$ with $r'$ being its spatial coordinate in the absence of metric perturbations, and $\delta \tau$ is the clock rate change, corresponding to the Einstein time delay. 
Keeping the linear terms, one can obtain the total time delay:
\begin{equation}
    \begin{split}
        c\delta T_{\gamma} \equiv cT_{\gamma} - 2L = &\underbrace{c\delta\tau}_\text{Einstein} + \underbrace{2r_\text{\rm M}\left(t+\frac{L}{c},L\right)-r_\text{\rm M}\left(t,0\right)-r_\text{\rm M}\left(t+\frac{2L}{c},0\right)}_\text{Doppler} \\
        &+ \underbrace{\frac{1}{2}\int_0^L dr \mathcal{H}^\text{\rm out}\left(t+\frac{r}{c},r\right) - \frac{1}{2}\int_L^0 dr \mathcal{H}^\text{\rm in}\left(t+\frac{2L-r}{c},r\right)}_\text{Shapiro}.
    \end{split}
    \label{eqn:Dopp_Shap}
\end{equation}
The second line shows the gravitational redshift experienced by the photon between two unperturbed mirrors, which corresponds to the definition of the Shapiro effect in the literature \cite{Shapiro1964,Siegel:2007fz}.  The contribution from mirror motion, corresponding to the Doppler effect, is the differential position shift between the two mirrors during one photon roundtrip obtained by solving the mirror geodesic equations, which, to leading order, can be written as~\cite{Baghram:2011is}
\begin{equation}
	r_\text{\rm M}(t,r) \approx \frac{1}{2}\int^t dt'\int^{t'}dt'' \partial_r \mathcal{H}_0(t'',r).
\end{equation}
The choice of initial conditions in the above time integrals can be subtle depending on the gauge choice, so it is more convenient to work with the mirror acceleration instead. The Doppler strain, which is a displacement quantity, is related to the mirror acceleration, $\tilde{a}(f)$, in the frequency domain via $|\tilde{h}_{\Doppler}(f)|^2\sim (2\pi f)^{-4}L^{-2}|\tilde{a}(f)|^2$~\cite{Fedderke:2021kuy}, circumventing the need to specify the mirrors' initial conditions. This is to be contrasted with the treatment of the Doppler effect in PTAs, where the accelerations of the pulsars / Earth are explicitly integrated over time to obtain the shift in the pulsar phase~\cite{Dror:2019twh, Ramani:2020hdo, Lee:2020wfn, Lee_2021}.

The total time delay in Eq.~\eqref{eqn:Dopp_Shap} is shown to be invariant under general gauge transformations \cite{Li:2022mvy}. Equipped with this well-defined observable, we specialize to the case of a transiting DM and compute each effect individually.

To compute the individual contributions, one has to fix a gauge. We work with the harmonic gauge, where the metric perturbation due to a nonrelativistic ($v\ll c$) point particle is given by~\cite{Carroll_2003} %
\begin{equation}\label{eqn:harmonic_gauge}
	ds^2 = -\left(1+\frac{2\Phi(t,\vec{x})}{c^2}\right)c^2dt^2+\left(1-\frac{2\Phi(t,\vec{x})}{c^2}\right)(dx^2+dy^2+dz^2) \, ,
\end{equation} 
where $\Phi$ is the DM Newtonian potential
\begin{equation}\label{eqn:DM_potential}
	\Phi(t,\vec{x}) = -\frac{GM}{|\vec{x}-\vec{r}_{\DM}(t)|} \, ,
\end{equation} 
with $\vec{r}_{\DM}(t)$ being the DM trajectory. Assuming a straight-line motion, we can completely specify $\vec{r}_{\DM}(t)$ using six phase space parameters $\{\vec{r}_0,\vec{v}\}$,
\begin{equation}\label{eqn:DM_trajectory}
	\vec{r}_{\DM}(t) = \vec{r}_0+\vec{v}t \, ,
\end{equation} 
where $\vec{r}_0$ and $\vec{v}$ are the three-dimensional DM initial position and velocity. While Eq.~\eqref{eqn:DM_trajectory} is intuitive, it is inconvenient to use in practice, since it does not explicitly show the DM time of closest approach, when the signal is maximized. For DM signals induced on a spatial point of the detector ({\it i.e.} Doppler effect and Einstein delay), a more convenient parametrization of Eq.~\eqref{eqn:DM_potential} is~\cite{Dror:2019twh}
\begin{equation}\label{eqn:DM_trajectory_b}
    \vec{r}_{\DM}(t) = \vec{b}+\vec{v}(t-t_0) \, ,
\end{equation} 
where $\vec{b}$ is the impact parameter, and $t_0$ can be interpreted as the DM ``arrival time". Note that $\vec{b}$ is constrained to be perpendicular to $\vec{v}$, hence the total number of phase space parameters $\{\vec{b},\vec{v},t_0\}$ is still six, and for an experiment with total observation time $T$, only DM with arrival time within the range $-T/2\leq t_0\leq T/2$ can be feasibly detected. One can rewrite Eq.~\eqref{eqn:DM_trajectory_b} as $\vec{r}_{\DM}(t) = b(\unit{b}+\eta\unit{v})$, where $\eta\equiv(t-t_0)/\tau$ is a dimensionless time parameter and $\tau\equiv b/v$.

Finally, if the DM signal depends on the closest distance between $\vec{r}_{\DM}$ and the experiment baseline ({\it i.e.}, Shapiro delay in the small impact parameter limit; see Sec.~\ref{subsec:shapiro}), assumed to be aligned in $\unit{n}$, then the most convenient parametrization of Eq.~\eqref{eqn:DM_trajectory} is
\begin{equation}\label{eqn:DM_trajectory_b_perp}
	\vec{r}_{\DM}(t) = \vec{b}_{\perp}+b_{\parallel}\unit{n}+\vec{v}(t-t_{0,\perp}) \, ,
\end{equation} 
where $\vec{b}_{\perp}$ and $b_{\parallel}$ are the perpendicular and parallel impact parameter respectively, and $t_{0,\perp}$ is the time when the DM reaches $\vec{b}_{\perp}$. The phase space parameters are $\{\vec{b}_{\perp},b_{\parallel},\vec{v},t_{0,\perp}\}$, where $b_{\perp}$ is constrained to be perpendicular to both $\unit{n}$ and $\unit{v}$, giving again a total of six independent parameters, as expected. The perpendicular DM distance is $r_{\perp}(t)\equiv|\vec{r}_{\DM}(t)\times\unit{n}|=b_{\perp}\sqrt{1+\eta_{\perp}^2}$, where $\eta_{\perp}\equiv(t-t_{0,\perp})/\tau_{\perp}$, $\tau_{\perp}\equiv b_{\perp}/v_{\perp}$ and $v_{\perp}\equiv |\vec{v}\times\unit{n}|$.

\subsection{Doppler Effect}
\label{subsec:doppler}

We start by studying the Doppler effect, focusing on a one-arm GW detector. The Doppler effect is often the only component of transiting DM signals considered in the literature, such as Refs.~\cite{Hall2016, Baum2022}, since it is the most dominant contribution in most mass ranges.  When unperturbed, suppose the two mirrors are located at $\vec{r}_{M_1}$ (an inner mirror close to the beamsplitter) and $\vec{r}_{M_2}$ (an exterior mirror at the edge of the arm), which are separated by a distance of $\vec{r}_{M_2}-\vec{r}_{M_1}=L\unit{n}$. The laser measures the distance between the two free-falling mirrors, with trajectories given by $\vec{r}_{M_1}(t)$ and $\vec{r}_{M_2}(t)$ (with arguments), evaluated at times separated by the photon transverse time. 
\begin{equation}\label{eqn:doppler}
	h_{\Doppler}(t) = \frac{\unit{n}}{L}\cdot\left\{\left[\vec{r}_{M_2}\left(t+\frac{L}{c}\right)-\vec{r}_{M_1}(t)\right]-\left[\vec{r}_{M_1}\left(t+\frac{2L}{c}\right)-\vec{r}_{M_2}\left(t+\frac{L}{c}\right)\right]\right\} \, .
\end{equation} 
This corresponds to the mirror motion term in Eq.~\eqref{eqn:Dopp_Shap}. It is in fact more natural to consider the Doppler signal as the acceleration of the mirrors caused by the transiting DM. The strain, which is a displacement quantity, is related to the mirror acceleration along the interferometer arm, $a_{M_a}(t)=\unit{n}\cdot\frac{d^2}{dt^2}\vec{r}_{M_a}(t)$ for $a=1,2$, by
\begin{equation}\label{eqn:mirror_acceleration}
	\frac{d^2}{dt^2}h_{\Doppler}(t) = \frac{1}{L}\left\{\left[a_{M_2}\left(t+\frac{L}{c}\right)-a_{M_1}(t)\right]-\left[a_{M_1}\left(t+\frac{2L}{c}\right)-a_{M_2}\left(t+\frac{L}{c}\right)\right]\right\} \, .
\end{equation} 
In the Newtonian limit, it is clear that the mirror accelerations are simply given by the gravitational potential from the DM. Alternatively, to more explicitly relate to the gauge invariant formalism developed in Ref.~\cite{Li:2022mvy}, one can also derive the mirror acceleration using the metric perturbation in Eq.~\eqref{eqn:harmonic_gauge}, which is a standard general relativity calculation that we briefly review. The mirrors free fall in accordance with the geodesic equation parametrized by the coordinate time, which is $\frac{d^2}{dt^2}r^{\mu}_{M_a}(t) + \Gamma^{\mu}_{\rho\sigma}[dr^{\rho}_{M_a}(t)/dt][dr^{\sigma}_{M_a}(t)/dt]=0$. For the metric in Eq.~\eqref{eqn:harmonic_gauge}, when the source is moving slowly ($v\ll c$), the Christoffel symbols are $\Gamma^0_{0i}=\Gamma^i_{00}=\partial_i \Phi/c^2$ and $\Gamma^i_{jk}=(\delta_{jk}\partial_i\Phi-\delta_{ik}\partial_j\Phi-\delta_{ij}\partial_k\Phi)/c^2$~\cite{Carroll_2003}. In the limit where the mirror is moving very slowly ($\dot{r}_{M_a}\ll c$), the leading order geodesic equation is $(d^2/dt^2)r^{i}_{M_a}(t) + c^2\Gamma^{i}_{00}=0$,
and thus the mirror acceleration is given by the gradient of the potential
\begin{equation}\label{eqn:acceleration_from_geodesic}
	a_{M_a}(t) = -\frac{GM}{\Delta r_{M_a}^2(t)}\Delta\unit{r}_{M_a}(t)\cdot\unit{n} \, ,
\end{equation} 
where we define the distance between the mirrors and the DM, $\Delta \vec{r}_{M_a}(t)\equiv\vec{r}_{M_a}-\vec{r}_{\DM}(t)$. This is of course the gravitational force that the DM exerts on the mirrors. We now take the DM trajectory in Eq.~\eqref{eqn:DM_trajectory_b} choosing the unperturbed beamsplitter location as the coordinate origin. Then the acceleration of the first mirror in Eq.~\eqref{eqn:acceleration_from_geodesic} is given by $a_{M_1}(t) = -\unit{n}\cdot(GM/b^2)(\hat{\vec{b}}+\eta\hat{\vec{v}})/(1+\eta^2)^{3/2}$
with the Fourier transform
\begin{align}\label{eqn:acceleration_M1_fourier}
	\tilde{a}_{M_1}(f) = -\frac{GM}{bv}e^{-2\pi ift_0}\tilde{s}_{M_1}\left(\frac{f}{f_{\tau}}\right)\quad 
{\rm with}\quad
 \tilde{s}_{M_1}(x) \equiv 2x\left[K_1(x)\unit{b}-iK_0(x)\unit{v}\right]\cdot\unit{n} \, ,
\end{align} 
where we separated the magnitude and the signal shape for clarity. The Doppler acceleration has a sharp peak at $\eta=0$ in real space, corresponding to the DM time of arrival as expected. In Fourier space, the Doppler acceleration has a weak log dependence on the frequency for $f<f_{\tau}$, but quickly drops to zero when $f>f_{\tau}$. This behavior is in fact general for all three types of signals, as we will see shortly.  

The acceleration of the second mirror can be computed using Eq.~\eqref{eqn:DM_trajectory}, Eq.~\eqref{eqn:acceleration_from_geodesic} and $\Delta\vec{r}_{M_2}(t)=\Delta\vec{r}_{M_1}(t)+L\unit{n}$,
\begin{equation}\label{eqn:acceleration_M2_parameterize}
	a_{M_2}(t) \approx \begin{dcases*}
		-\frac{GM}{L^2}, & if $b\ll L$\\
		a_{M_1}(t)+\frac{GML}{b^3}\left[\frac{3(\unit{b}\cdot\unit{n}+\eta\unit{v}\cdot\unit{n})^2}{(1+\eta^2)^{5/2}}-\frac{1}{(1+\eta^2)^{3/2}}\right], & if $b\gg L$
	\end{dcases*} \, .
\end{equation} 
The Fourier transform is given by
\begin{equation}\label{eqn:acceleration_M2_fourier}
	\tilde{a}_{M_2}(f) \approx \begin{dcases*}
		0, & if $b\ll L$\\
		\tilde{a}_{M_1}(f)+\left(\frac{L}{b}\right)\tilde{a}_{\tidal}(f), & if $b\gg L$
	\end{dcases*} \, ,
\end{equation} 
where we 
have approximated $ \delta(f) \sim 0$ and
\begin{align}\label{eqn:a_tidal}
	\tilde{a}_{\tidal}(f) &= \frac{GM}{bv}e^{-2\pi ift_0}\tilde{s}_{\tidal}\left(\frac{f}{f_{\tau}}\right) \nonumber \\
    \tilde{s}_{\tidal}(x) &\equiv 2x \left\{\left[{(\unit{b}\cdot\unit{n})}^2-{(\unit{v}\cdot\unit{n})}^2\right]xK_0(x)+\left[\left(2{(\unit{b}\cdot\unit{n})}^2+{(\unit{v}\cdot\unit{n})}^2-1\right)-2i(\unit{b}\cdot\unit{n})(\unit{v}\cdot\unit{n})x\right]K_1(x)\right\}\,. 
\end{align} 
The total Doppler effect can be computed by Fourier transforming Eq.~\eqref{eqn:mirror_acceleration} and, using Eq.~\eqref{eqn:acceleration_M2_fourier}, we have
\begin{align}\label{eqn:doppler_Fourier}
	\tilde{h}_{\Doppler}(f) 
	&\approx -\frac{2}{(2\pi f)^2L}e^{-if/2f_{\FSR}}\begin{dcases*}
	\cos\left(\frac{f}{2f_{\FSR}}\right)\tilde{a}_{M_1}(f), & if $b\ll L$ \\
	\underbrace{2\sin^2\left(\frac{f}{4f_{\FSR}}\right)\tilde{a}_{M_1}(f)}_{\text{Q-suppressed}}-\underbrace{\left(\frac{L}{b}\right)\tilde{a}_{\tidal}(f)}_{\text{tidal-suppressed}}, & if $b\gg L$ 
	\end{dcases*} \, . 
\end{align} 
Equation~\eqref{eqn:doppler_Fourier} is illuminating, as it shows that we can safely ignore the contribution from the second mirror if $b\ll L$ since the acceleration of the first mirror is much greater than the second mirror. However, if $b\gg L$, then both mirrors experience similar acceleration from the same transiting DM. This leads to a suppression factor of $(L/b)$, known as the tidal factor as alluded to in Sec.~\ref{sec:signal_description}, which is well studied in the literature of accelerometers~\cite{Jaeckel:2020mqa,Baum2022}. However, an additional piece of the power spectrum is not suppressed by the tidal factor, but arises from the fact that a laser interferometer measures the differential acceleration between the mirrors at a slightly different time. Defining $Q_f \equiv f_{\FSR}/f$, the Doppler signal is thus
\begin{equation}\label{eqn:doppler_Fourier_psd}
	|\tilde{h}_{\Doppler}(f)|^2 \approx \left(\frac{8GML}{c^2bv}\right)^2\left(\frac{f_{\FSR}}{f}\right)^4 \begin{dcases*}
		\cos^2\left(\frac{f}{2f_{\FSR}}\right)\left|\tilde{s}_{M_1}\left(\frac{f}{f_{\tau}}\right)\right|^2, & if $b\ll L$ \\
		4\sin^4\left(\frac{f}{4f_{\FSR}}\right)\left|\tilde{s}_{M_1}\left(\frac{f}{f_{\tau}}\right)\right|^2, & if $b\gg L$ and $1/Q_f \gg L/b$ \\
		\left(\frac{L}{b}\right)^2\left|\tilde{s}_{\tidal}\left(\frac{f}{f_{\tau}}\right)\right|^2, & if $b\gg L$ and $1/Q_f \ll L/b$
	\end{dcases*} \, ,
\end{equation} 
where we assume that either the $Q$-suppressed or the tidal-suppressed term dominates when $b\gg L$. Effects from the finite photon travel time in laser interferometers have been considered previously in the literature of gravitational wave \cite{Larson:1999, Cornish:2002, Rakhmanov:2004, Rakhmanov:2008} and ultralight DM \cite{Morisaki:2018, Morisaki2021}. However, previous studies on macroscopic DM have generally neglected this effect by treating the interferometer as a simple accelerometer, which amounts to dropping the sines and cosines of $\sim(f/f_{\FSR})$, the second term of Eq.~\eqref{eqn:doppler_Fourier_psd}, and estimating $\tilde{s}_{\tidal}(f/f_{\tau})\approx \tilde{s}_{M_1}(f/f_{\tau})$. This treatment is well justified for GW detectors that utilize FP cavities with $Q\gg 1$ such as LIGO, but does not apply to other laser interferometers such as Holometer and GQuEST.

The average signal shape can be computed by taking the amplitude squared of Eq.~\eqref{eqn:acceleration_M1_fourier} and Eq.~\eqref{eqn:a_tidal} while substituting the angular factors derived in App.~\ref{app:angular_factors} (see Eq.~\eqref{eqn:angular_mean} and Eq.~\eqref{eqn:angular_four}),
\begin{align}\label{eqn:doppler_s_averaged}
	\left\langle|\tilde{s}_{M_1}(x)|^2\right\rangle &= \frac{4}{3}x^2\left[K_0^2(x)+K_1^2(x)\right] 
 \approx \frac{4}{3} \begin{dcases*}
		1, & if $x\ll 1$ \\
		\pi xe^{-2x}, & if $x\gg 1$
	\end{dcases*} \nonumber \\
	\left\langle|\tilde{s}_{\tidal}(x)|^2\right\rangle &= \sum_{i,j,k}c_{ij,k}x^kK_i(x)K_j(x) 
 \approx \frac{16}{15} \begin{dcases*}
		1, & if $x\ll 1$ \\
		\pi x^3e^{-2x}, & if $x\gg 1$
	\end{dcases*} \, ,
\end{align} 
where $c_{ij,k}$ are some $\mathcal{O}(1)$ coefficients. We observe that $ \left\langle|\tilde{s}_{\tidal}(x)|^2\right\rangle\sim(4/5)\left\langle|\tilde{s}_{M_1}(x)|^2\right\rangle$. The spectral shapes of the tidal-suppressed piece and the $Q$-suppressed piece are in fact very similar to each other (constant until $f_{\tau}$ and then exponential decay), but are suppressed by factors of different physical origins. To set an upper limit on the DM fraction, we compute the 10$^{\text{th}}$ percentile $|\tilde{h}_{\Doppler}(f)|^2$ using Eq.~\eqref{eqn:doppler_Fourier_psd} with the impact parameter given by Eq.~\eqref{eqn:b_min}, while taking the mean value of the angular factors and $v$. In the limit where $b\ll L$, one recovers Eq.~\eqref{eqn:simplified_spectrum}. On the other hand, if the interferometer has two arms separated by an angle of $\Delta\theta$, then we replace the angular factors in Eq.~\eqref{eqn:doppler_s_averaged} with the two-arm angular factors in Eq.~\eqref{eqn:angular_twoarn}, while removing the $\unit{n}$-independent term in Eq.~\eqref{eqn:a_tidal}, giving, up to $\mathcal{O}(1)$ factors, a factor of $4\sin^2(\Delta \theta/2)$ in Eq.~\eqref{eqn:doppler_Fourier_psd}.

\subsection{Shapiro Delay}
\label{subsec:shapiro}

The Shapiro delay has been studied extensively for transiting DM signals in PTAs~\cite{Siegel:2007fz,Dror:2019twh,Ramani:2020hdo}. A pulsar located at a distance of $z_0$ from Earth has a long baseline of $L\sim$ kpc, which is greater than the DM impact parameter even for DM as heavy as 10 $M_{\odot}$. Hence most PTA works compute the Shapiro signal assuming $z_0>b$, in which case it has been shown that the relevant impact parameter is defined relative to the line of sight between Earth and the pulsar, {\it i.e.} $b_{\perp}$. For laser interferometers, however, the baseline $L$ can, in general, be smaller than the DM impact parameter, $b$. In this section, we show that if $b>L$, then the relevant impact parameter is, in fact, $b$ as the length scale of the detector becomes negligible and the detector becomes ``pointlike", which is consistent with Ref.~\cite{Siegel:2007fz}. In the opposite limit where $b<L$, an interferometer becomes similar to the pulsar-Earth system, and the relevant impact parameter is $b_{\perp}$.  

We choose the midpoint of the interferometer arm as the coordinate origin in  Eq.~\eqref{eqn:DM_trajectory_b_perp} . The total Shapiro delay is given by the change in the proper length of the interferometer arm, measured over a photon roundtrip
\begin{equation}\label{eqn:Shapiro}
	h_{\Shapiro}(t) = \frac{1}{L}\left[\Delta l\left(t+\frac{L}{2c}\right)+\Delta l\left(t+\frac{3L}{2c}\right)\right] \, ,
\end{equation} 
where $\Delta l(t)$ is the shift in the proper arm length, measured by a photon that passes through the midpoint of the arm, $\vec{r}_{\mid}$, at time $t$, {\it i.e.} $\Delta l(t) = \frac{1}{2}\int_{-L/2}^{L/2}dz\,h_{ij}\left(t+\frac{z}{c},\vec{r}_{\mid}+z\unit{n}\right)n^in^j$. Under the DM gravitational field in Eq.~\eqref{eqn:harmonic_gauge} and Eq.~\eqref{eqn:DM_potential}, and assuming that DM moves in a straight line with a constant velocity according to Eq.~\eqref{eqn:DM_trajectory}, the motion of $\vec{r}_{\DM}(t)$ within a one-way photon travel time can be simply treated as $\vec{r}_{\DM}(t+(z/c))=\vec{r}_{\DM}(t)+(z/c)\vec{v}$, and hence~\cite{Siegel:2007fz}
\begin{equation}\label{eqn:proper_length_shift_4}
	\Delta l(t) = \frac{GM}{c^2}\int_{-L/2}^{L/2}dz\frac{1}{\sqrt{(z-r_{\parallel})^2+r_{\perp}^2}}\, ,
\end{equation} 
where we defined the parallel and perpendicular distance of DM, relative to the arm midpoint, $r_{\perp} \equiv |[\vec{r}_{\DM}(t)-\vec{r}_{\mid}]\times \unit{n}|$ and $r_{\parallel} \equiv [\vec{r}_{\DM}(t)-\vec{r}_{\mid}]\cdot\unit{n}$, and used the nonrelativistic limit $\unit{n}-(v/c)\unit{v}\approx\unit{n}$. The integral in Eq.~\eqref{eqn:proper_length_shift_4} can now be computed analytically~\cite{Siegel:2007fz}
\begin{equation}\label{eqn:proper_length_integrated}
	\Delta l(t) = \frac{GM}{c^2}\log\left(\frac{r_{\parallel}+(L/2)+\sqrt{r_{\perp}^2+[r_{\parallel}+(L/2)]^2}}{r_{\parallel}-(L/2)+\sqrt{r_{\perp}^2+[r_{\parallel}-(L/2)]^2}}\right) 
 \approx \frac{GM}{c^2}\begin{dcases*}
		\log\left(\frac{L^2}{r_{\perp}^2}\right), & if $\sqrt{r_{\perp}^2+r_{\parallel}^2}\lesssim L/2$ \\
		\frac{L}{\sqrt{r_{\perp}^2+r_{\parallel}^2}}, & if $\sqrt{r_{\perp}^2+r_{\parallel}^2}\gtrsim L/2$
	\end{dcases*} \, ,
\end{equation} 
taking a simpler form in the two different limits.
We observe that when $b\gtrsim L/2$, then the Shapiro delay depends on the magnitude of $\vec{b}$, similar to the Doppler effect. However when $b\lesssim L/2$, the effect only depends on the perpendicular component and has a weak log boost from the small distance (as opposed to the Doppler case). Using the parametrization in Eqs.~\eqref{eqn:DM_trajectory_b}-\eqref{eqn:DM_trajectory_b_perp} for the upper and lower entries of Eq.~\eqref{eqn:proper_length_integrated} and performing a Fourier transform, we find~\footnote{Useful Fourier transform integral: $\int_{-\infty}^{\infty} dx e^{-ikx} \log\left(\frac{1}{\alpha x^2+\beta x+\gamma}\right) = \frac{2\pi}{|k|}e^{ik\frac{\beta}{2\alpha}}e^{-\frac{\sqrt{-\Delta}}{2\alpha}|k|}$, where ${\alpha,\beta\,\gamma}\in\mathbb{R}$, $\alpha>0$, $\gamma>0$, and $\Delta\equiv\beta^2-4\alpha\gamma<0$. We dropped all delta functions as usual.}
\begin{equation}\label{eqn:proper_length_limits_Fourier}
	\Delta \tilde{l}(f) \approx \frac{GM}{c^2}\begin{dcases*}
		\frac{1}{f}e^{-2\pi ift_{0,\perp}}e^{-f/f_{\tau_{\perp}}}, & if $b\lesssim L/2$ \\
		\frac{2L}{v} e^{-2\pi ift_0}K_0\left(\frac{f}{f_{\tau}}\right), & if $b\gtrsim L/2$
	\end{dcases*} \, .
\end{equation} 
We then Fourier transform the total Shapiro delay in Eq.~\eqref{eqn:Shapiro}:
\begin{align}\label{eqn:Shapiro_fourier_magnitude}
	|\tilde{h}_{\Shapiro}(f)|^2 = \left(\frac{8\pi GM}{c^3}\right)^2\cos^2\left(\frac{f}{4f_{\FSR}}\right)\begin{dcases*}
		\left(\frac{f_{\FSR}}{f}\right)^2e^{-2f/f_{\tau_{\perp}}}, & if $b\lesssim L/2$ \\
		\left(\frac{c}{2\pi v}\right)^2K_0^2\left(\frac{f}{f_{\tau}}\right), & if $b\gtrsim L/2$
	\end{dcases*} \, .
\end{align} 
We emphasize that Eq.~\eqref{eqn:Shapiro_fourier_magnitude} should be read with caution. The two entries of Eq.~\eqref{eqn:Shapiro_fourier_magnitude} are decided by the relative magnitude between $b$ and $L/2$. If $b\gtrsim L/2$, then the Shapiro signal is cut off at the frequency corresponding to $b$. Otherwise if $b\lesssim L/2$, then the Shapiro spectrum is suppressed by factors of $v/c$ compared to the Doppler spectrum in Eq.~\eqref{eqn:doppler_Fourier_psd}, and is cut off at the frequency corresponding to $b_{\perp}$. %

In a two-arm interferometer system with an $\mathcal{O}(1)$ (in radians) arm separation angle, the total Shapiro strain is the difference between individual arm strains. If $b\ll L/2$, then the Shapiro delay for one of the arms should be much stronger than that of the second arm (it is statistically unlikely that the DM with the smallest $b_{\perp,1}$ for one arm also has a comparably small $b_{\perp,2}$ relative to the second arm unless the angle between the two arms is very small, which is in general not true for any realistic GW detector). Otherwise, if $b\gg L/2$, then the two interferometer arms effectively become two point detectors oriented toward directions $\unit{n}_1$ and $\unit{n}_2$, and the total strain suffers a tidal suppression factor of $L/b$, similar to the Doppler effect in Eq.~\eqref{eqn:acceleration_M2_parameterize}. Using Eq.~\eqref{eqn:proper_length_integrated}, we find
\begin{equation}\label{eqn:proper_length_two_arm}
	\Delta l(t,\unit{n}_1)-\Delta l(t,\unit{n}_2) 
 \approx \frac{GM}{c^2}\begin{dcases*}
		\log\left(\frac{L^2}{|\vec{r}_{\DM}(t)\times\unit{n}_1|^2}\right), & if $b\lesssim L/2$ \\
		\frac{L^2}{2r_{\DM}^2(t)}\unit{r}_{\DM}(t)\cdot(\unit{n}_1-\unit{n}_2), & if $b\gtrsim L/2$
	\end{dcases*} \, .
\end{equation} 
Taking the Fourier transform of Eq.~\eqref{eqn:proper_length_two_arm} using Eq.~\eqref{eqn:DM_trajectory_b_perp} and Eq.~\eqref{eqn:Shapiro}, the total Shapiro strain is
\begin{align}\label{eqn:Shapiro_fourier_two_arm}
	|\tilde{h}_{\Shapiro}(f)|^2 = \left(\frac{8\pi GM}{c^3}\right)^2\cos^2\left(\frac{f}{4f_{\FSR}}\right)\begin{dcases*}
		\left(\frac{f_{\FSR}}{f}\right)^2e^{-2f/f_{\tau_{\perp}}}, & if $b\lesssim L/2$ \\
		\left(\frac{L}{8b}\right)^2\left(\frac{c}{2\pi v}\right)^2\left|\tilde{s}_{M_1}\left(\frac{f}{f_{\tau}}\right)\right|^2, & if $b\gtrsim L/2$
	\end{dcases*} \, ,
\end{align} 
where $\tilde{s}_{M_1}(x)$ is defined in Eq.~\eqref{eqn:acceleration_M1_fourier}, but with $\unit{n}$ replaced by $\unit{n}_1-\unit{n}_2$. Comparing Eq.~\eqref{eqn:Shapiro_fourier_two_arm} with Eq.~\eqref{eqn:Shapiro_fourier_magnitude}, we see that the Shapiro spectrum for a two-arm interferometer is identical to that of a one-arm detector for $b\lesssim L/2$, but picks up a suppression factor of $4\sin^2(\Delta \theta/2)(L/8b)^2$ when $b\gtrsim L/2$.

\subsection{Einstein Delay}
\label{subsec:einstein}

The gravitational effect due to the Einstein delay is given by the difference of the clock proper time $\tau$ at the beamsplitter over a photon roundtrip time
\begin{equation}\label{eqn:Enistein}
	h_{\Einstein}(t) = \frac{c}{L}\left[\tau\left(t+\frac{2L}{c}\right)-\tau(t)\right] \, .
\end{equation} 
The proper and coordinate times are related by $d\tau(t)/dt=1-(1/2)h_{00}$, so that using Eq.~\eqref{eqn:harmonic_gauge} and Eq.~\eqref{eqn:Enistein}, we write
\begin{equation}\label{eqn:Einstein_time_derivative}
	\frac{d}{dt}h_{\Einstein}(t) = \frac{1}{cL}\left[\Phi\left(t+\frac{2L}{c}\right)-\Phi(t)\right] \, ,
\end{equation} 
where the DM gravitational potential $\Phi$ is evaluated at the beamsplitter. Putting the DM trajectory in Eq.~\eqref{eqn:DM_trajectory} with the beamsplitter location chosen as the coordinate origin into the potential, one finds $\Phi(t) = -(GM/b)(1+\eta^2)^{-1/2}$ with the Fourier transform $\tilde{\Phi}(f) = -(2GM/v)e^{-2\pi ift_0}K_0(f/f_{\tau})$, giving the Einstein strain
\begin{equation}\label{eqn:Einstein_Fourier_magnitude}
	|\tilde{h}_{\Einstein}(f)|^2 = \left(\frac{8GM}{c^2v}\right)^2\left(\frac{f_{\FSR}}{f}\right)^2\sin^2\left(\frac{f}{2f_{\FSR}}\right)K_0^2\left(\frac{f}{f_{\tau}}\right) \, .
\end{equation} 
Note that the impact parameter only enters the spectrum through the peak frequency $f_{\peak}$, but not the amplitude.

If the interferometer has two arms, then the Einstein delay contribution cancels between the two interferometer arms, and thus the effect vanishes.

\section{Stochastic Signal}
\label{sec:stochastic}

In the small DM mass limit, it is possible that each individual DM is not sufficient to produce a sizable signal, but the collective effect due to all DM passing by the detector might be large enough to be measured. In this limit, DM behaves collectively like a stochastic background. The total strain $h(t)$ is given by summing over strains $h_a(t)$ from all individual DM:
\begin{equation}\label{eqn:total_stochastictrain}
	h(t) = \sum_a h_a(t) \, .
\end{equation} 
Correlations from the stochastic DM field have been previously studied in Ref.~\cite{Ramani:2020hdo} in the context of PTAs.

\subsection{Doppler Effect}\label{subsec:doppler_stochastic}

For a given $\unit{v}$, the differential volume of an element in a cylinder is $dV=v b db d\varphi dt_0$, where $\varphi$ is the polar angle of $\vec{b}$ on the plane perpendicular to $\vec{v}$. 
We have assumed monochronic DM masses. Using the parametrization in Eq.~\eqref{eqn:DM_trajectory_b}, the autocorrelation function of $\tilde{h}_{\Doppler}(f)$ is then given by integrating over the volume
\begin{equation}\label{eqn:sum_af}
	\langle\tilde{h}(f)\tilde{h}^*(f')\rangle = \frac{n}{4\pi}\int_{b_{\min}}^{\infty} bdb \int_0^{2\pi}d\varphi \int_{-T/2}^{T/2}dt_0 \int_0^{v_{\mathrm{esc}}} vf_v(v)dv \int_0^{\pi}\sin\theta d\theta\int_0^{2\pi}d\phi\,\tilde{h}(f)\tilde{h}^*(f') \, ,
\end{equation} 
where $f_v(v)$ is the Maxwell-Boltzmann distribution for the velocity, and $\theta$ and $\phi$ are the polar and azimuthal angles for $\mathbf{v}$ respectively. The factor of $1/(4\pi)$ comes from normalization of the angular integration over $\theta$ and $\phi$. Note that we also set the lower limit of the integral over $b$ to $b_{\min}$ as defined in the RHS of Eq.~\eqref{eqn:b_min}. The integral is formally divergent if we allow $b\to 0$, which is a case of statistical outliers skewing the mean of a distribution. Following the treatment of Ref.~\cite{Ramani:2020hdo} in the context of PTAs, the divergence can be regulated by truncating the integral at the $90^{\th}$ percentile of minimum impact parameter among DM particles, which is insensitive to statistical outliers.

Since the DM trajectory in Eq.~\eqref{eqn:DM_trajectory} is a function of $t-t_0$, the strain in Fourier space $\tilde{h}(f)$ can  only depend on $t_0$ through a phase factor $\exp(-2\pi ift_0)$. Integrating over $t_0$ thus evaluates to a delta function in $f-f'$ in the limit when $fT\gg 1$: $\langle\tilde{h}_{\Doppler}(f)\tilde{h}_{\Doppler}(f')\rangle = S_{\Doppler}(f)\delta(f-f') $, indicating that the stochastic signal is stationary when the observation time is sufficiently large. The strain power spectrum reads as
\begin{equation}\label{eqn:af_delta}
	S_{\Doppler}(f) = \frac{n}{4\pi}\int_{b_{\min}}^{\infty} bdb \int_0^{2\pi}d\varphi \int_0^{v_{\mathrm{esc}}} vf_v(v)dv \int_0^{\pi}\sin\theta d\theta\int_0^{2\pi}d\phi\,|\tilde{h}(f)|^2 \, .
\end{equation} 
Anticipating that the signal's dependence on the velocity is going to be weak, we set the velocity to $\bar{v}$, while integrating over angular factors of $\hat{\mathbf{b}}\cdot\hat{\mathbf{n}}$ and $\hat{\mathbf{v}}\cdot\hat{\mathbf{n}}$ using Eq.~\eqref{eqn:angular_mean}. We show the analytic form for the Doppler effect and extend the spectrum result for $b<L$ to $b\to\infty$, since we expect the detector to only be sensitive to $b<L$. Then integrating Eq.~\eqref{eqn:af_delta} with Eq.~\eqref{eqn:doppler_Fourier_psd}, we find~\footnote{Useful integral: $\int_a^{\infty}x[K^2_0(x)+K^2_1(x)]dx=aK_0(a)K_1(a)$ for $a>0$.}
\begin{align}\label{eqn:two_point_a_integral_2}
	S_{\Doppler}(f) 
	= B_{\Doppler}\left(\frac{f_{\FSR}}{f}\right)^4 \cos^2\left(\frac{f}{2f_{\FSR}}\right) \left(\frac{f}{f_{\tau_{\min}}}\right)K_0\left(\frac{f}{f_{\tau_{\min}}}\right)K_1\left(\frac{f}{f_{\tau_{\min}}}\right)\, ,
\end{align} 
where $f_{\tau_{\min}}\equiv v/(2\pi b_{\min})$, and 
\begin{align}\label{eqn:stochastic_PSD}
	B_{\Doppler} &\equiv \frac{128\pi L^2G^2M\rho_{\DM}f_{\DM}}{3c^4\bar{v}} 
 = 2\times 10^{-73}\,\Hz^{-1}\left(\frac{M}{\kg}\right)f_{\DM}\left(\frac{340\,\kms}{\bar{v}}\right)\left(\frac{L}{\km}\right)^2 \, .
\end{align} 
For stationary signals, considering the cross-correlation between two detectors and the optimal matched filtering, the stochastic SNR is given by 
\begin{equation}\label{eqn:SNR_stochastic}
	\mathrm{SNR}_{\mathrm{X,stoc}}^2 = 2T\int_0^{\infty}df\, \Gamma^2 (f)\,\frac{S_X^2(f)}{S_{n_1}(f)S_{n_2}(f)} \, ,
\end{equation} 
where $S_{n_{1,2}}(f)$ are the one-sided autocorrelated PSD of the two detectors $1,2$, $\Gamma(f)$ is the cross-correlation function across detectors, and $X$ can correspond to the Doppler, Shapiro, or Einstein effect. For simplicity, we assume $\Gamma(f) \sim 1$, {\it i.e.}~the two detectors are colocated and aligned without correlated noise, and $S_{n_{1}}(f) = S_{n_{2}}(f) = S_{n}(f)$. Note that if there is only one detector, then due to the random nature of both the signal and the noise, no matched filtering can be applied.

\begin{figure*}[t]
	\includegraphics[scale=0.65]{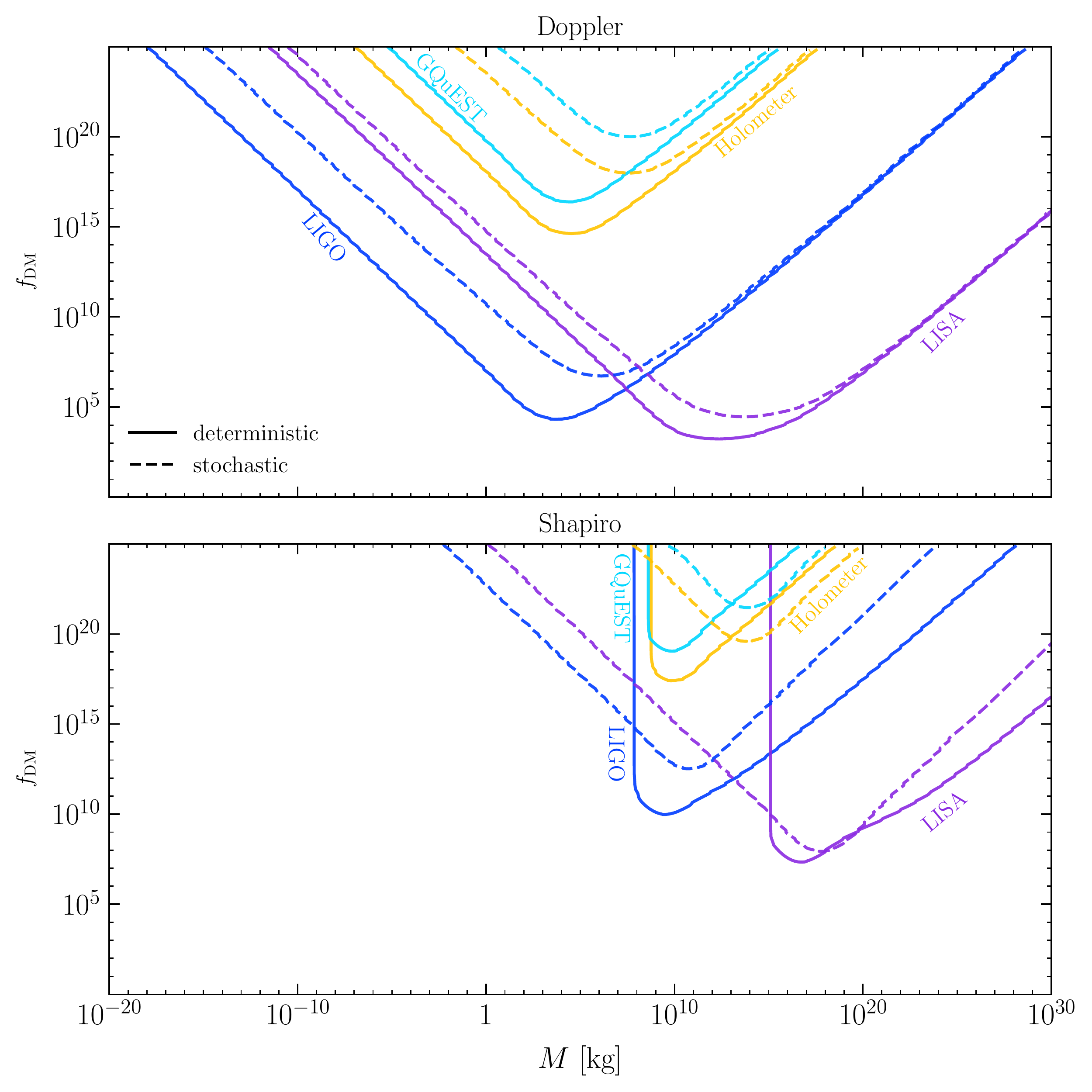} 
	\caption{Projected 90$^{\th}$-percentile upper limits on stochastic DM signals from LIGO, LISA, GQuEST, and Holometer, assuming $T=1$ year of observation time and local DM density $\rho_{\DM}=0.46$ GeV/$c^2$/cm$^3$. The limits are derived by setting the $10^{\th}$-percentile SNR defined in Eq.~\eqref{eqn:SNR_stochastic} to be 2. See Sec.~\ref{sec:sensitivity} for a description of the experiments.}\label{fig:stochastic}
\end{figure*}

In the first panel of~Fig.~\ref{fig:stochastic}, we show constraints on the DM fraction $f_{\rm DM}$ for LIGO, LISA, GQuEST, and Holometer, which all have two detectors. For the mass range and experimental parameters considered in this work, the Doppler stochastic reach is subdominant compared to the deterministic reach, which is consistent with the conclusion of Ref.~\cite{Hall2016}. This can be explicitly shown by estimating the SNR assuming both the signal and the noise are constant within the experiment's frequency window $\Delta f$; using~Eqs.~\eqref{eqn:doppler_Fourier_psd} and \eqref{eqn:two_point_a_integral_2}, one observes
\begin{equation}\label{eqn:SNR_deterministic_estimate}
   \frac{\rm SNR_{\rm stoc,\,\Doppler}}{(\rm SNR_{det,\,\Doppler})^2} \approx \frac{1}{\sqrt{T \Delta f}} \,
   \left. \frac{K_0(x) K_1(x)}{x [K_0(x)^2 + K_1(x)^2 ]} \right|_{x = f_{\rm peak}/f_{\tau}}  \approx -\frac{1}{\sqrt{T \Delta f}}  \log (f_{\rm peak}/f_{\tau}) \, ,
\end{equation}
in the limit $f_{\rm peak} \ll f_{\tau}$, showing that the stochastic constraint grows logarithmically for lower masses. As for all GW detectors considered in this work, including LIGO, LISA, GQuEST and Holometer, $T \Delta f \gg 1$, and hence the stochastic limit is subdominant for the whole mass range as shown in~Fig.~\ref{fig:stochastic}. Notice that for PTAs, for example, $T \Delta f \sim 1$ thus the stochastic constraint can take over at a relevant mass range \cite{Ramani:2020hdo}.

\subsection{Shapiro Delay}
The stochastic Shapiro delay can be derived in a similar manner. Recall from Sec.~\ref{subsec:shapiro} that in the $b\ll L$ limit, the impact parameter relative to the interferometer arm sets the size of the signal. The volume of a differential element is given by $dV=v_{\perp}db_{\perp}db_{\parallel}dt_{0,\perp}$~\cite{Ramani:2020hdo}. The stochastic Shapiro power spectrum is thus given by integrating the Shapiro strain in Eq.~\eqref{eqn:Shapiro_fourier_magnitude} with the parametrization in Eq.~\eqref{eqn:DM_trajectory_b_perp} over volume
\begin{equation}\label{eqn:Shapiro_stochastic_integral}
	S_{\Shapiro}(f) = \frac{n}{4\pi}\int_{0}^{\infty} db_{\perp} \int_{-L/2}^{L/2}db_{\parallel} \int_0^{v_{\mathrm{esc}}} v_{\perp}f_{v_{\perp}}(v_{\perp})dv \int_0^{\pi}\sin\theta d\theta\int_0^{2\pi}d\phi\,|\tilde{h}_{\Shapiro}(f)|^2 \, ,
\end{equation}
where we again assume that $\vec{v}$ points at a direction of $(\theta,\phi)$, $f_{v_{\perp}}(v_{\perp})$ is the perpendicular (relative to the interferometer arm) velocity distribution, and we performed the integral over $t_{0,\perp}$ assuming $fT\gg 1$. Anticipating that DM with $b_{\perp}<L/2$ will dominate the stochastic signal, we perform the integral in Eq.~\eqref{eqn:Shapiro_stochastic_integral} using the upper entry of Eq.~\eqref{eqn:Shapiro_fourier_magnitude} and find
\begin{align}\label{eqn:Shapiro_stochastic_spectrum}
	S_{\Shapiro}(f) 
	= B_{\Shapiro}\left(\frac{f_{\FSR}}{f}\right)^3 \cos^2\left(\frac{f}{4f_{\FSR}}\right)\left(1-e^{-(f/2f_{\FSR})(c/v_{\perp})}\right)\Theta(L-b_{\perp,\min}) \, ,
\end{align} 
where
\begin{align}\label{eqn:B_shapiro}
	B_{\Shapiro} &\equiv \frac{64\pi^2 L^2G^2M\rho_{\DM}f_{\DM}\bar{v}_{\perp}^2}{c^7} 
 = 8\times 10^{-82}\,\Hz^{-1}\left(\frac{M}{\kg}\right)f_{\DM}\left(\frac{\bar{v}_{\perp}}{270\,\kms}\right)^2\left(\frac{L}{\km}\right)^2 \, ,
\end{align} 
and we substituted the mean value, $\bar{v}_{\perp}$, of the velocity distribution, and introduced a cutoff requiring $b_{\perp,\min}<L$ to ensure that there is a nonzero number of DM with $b_{\perp}<L$.

The projected reach of the stochastic Shapiro signal is shown in the second panel of Fig.~\ref{fig:stochastic}, and is derived by setting the SNR in Eq.~\eqref{eqn:SNR_stochastic} with the spectrum in Eqs.~\eqref{eqn:Shapiro_stochastic_spectrum}-\eqref{eqn:B_shapiro} to be 2. Unlike the Doppler effect, the stochastic Shapiro signal can actually have better reach than the deterministic signal in the lower DM mass range, such as $M\lesssim 10^6$ kg for LIGO, for example. This can be traced to the fact that the Shapiro deterministic signal amplitude is independent of the DM mass once the DM impact parameter becomes less than $L$. However, the stochastic Shapiro spectrum scales linearly with the DM mass, as is evident in Eq.~\eqref{eqn:B_shapiro}, resulting in a larger SNR for the relevant mass range. 

The stochastic signal derived in this section in general agrees with the results in Ref.~\cite{Ramani:2020hdo} studied in the context of PTAs. In particular, the stochastic spectrum in Eq.~\eqref{eqn:stochastic_PSD} and that in Eq.~\eqref{eqn:Shapiro_stochastic_spectrum} have the same scaling relations of $M$, $f_{\DM}$, $\rho_{\DM}$, $\bar{v}$, $\bar{v}_{\perp}$ and $L$ (up to the definition of the observable) as the spectrum derived in Ref.~\cite{Ramani:2020hdo}. A notable difference is that Ref.~\cite{Ramani:2020hdo} presented the stochastic signal as a nonstationary process, with a power spectrum that is a function of both $f$ and $f'$, as opposed to the stationary signal we derived in this section, where the power spectrum is only a function of $f$, similar to a stochastic GW background~\cite{Christensen:2018iqi}. As discussed in Sec.~\ref{subsec:doppler_stochastic}, one can explicitly demonstrate that the stochastic DM signal is stationary by integrating over the DM arrival time, $t_0$ (Doppler) or $t_{0,\perp}$ (Shapiro), over the experimental time $T$. In the limit $fT\gg 1$, which holds for all GW detectors considered in this work with $T=1$ year, one finds that $\langle \tilde{h}(f)\tilde{h}(f')\rangle$ is proportional to $\delta(f-f')$, which is the definition of a stationary process. Physically, this demonstrates that for a sufficiently long observation time, DM can arrive at any time during the experiment, which is a uniformly random variable, and hence the signal produced is stationary in nature. %

\section{Gravitational wave experiments and noise curves}\label{sec:sensitivity}

\begin{figure*}
    \includegraphics[width=0.9\linewidth]{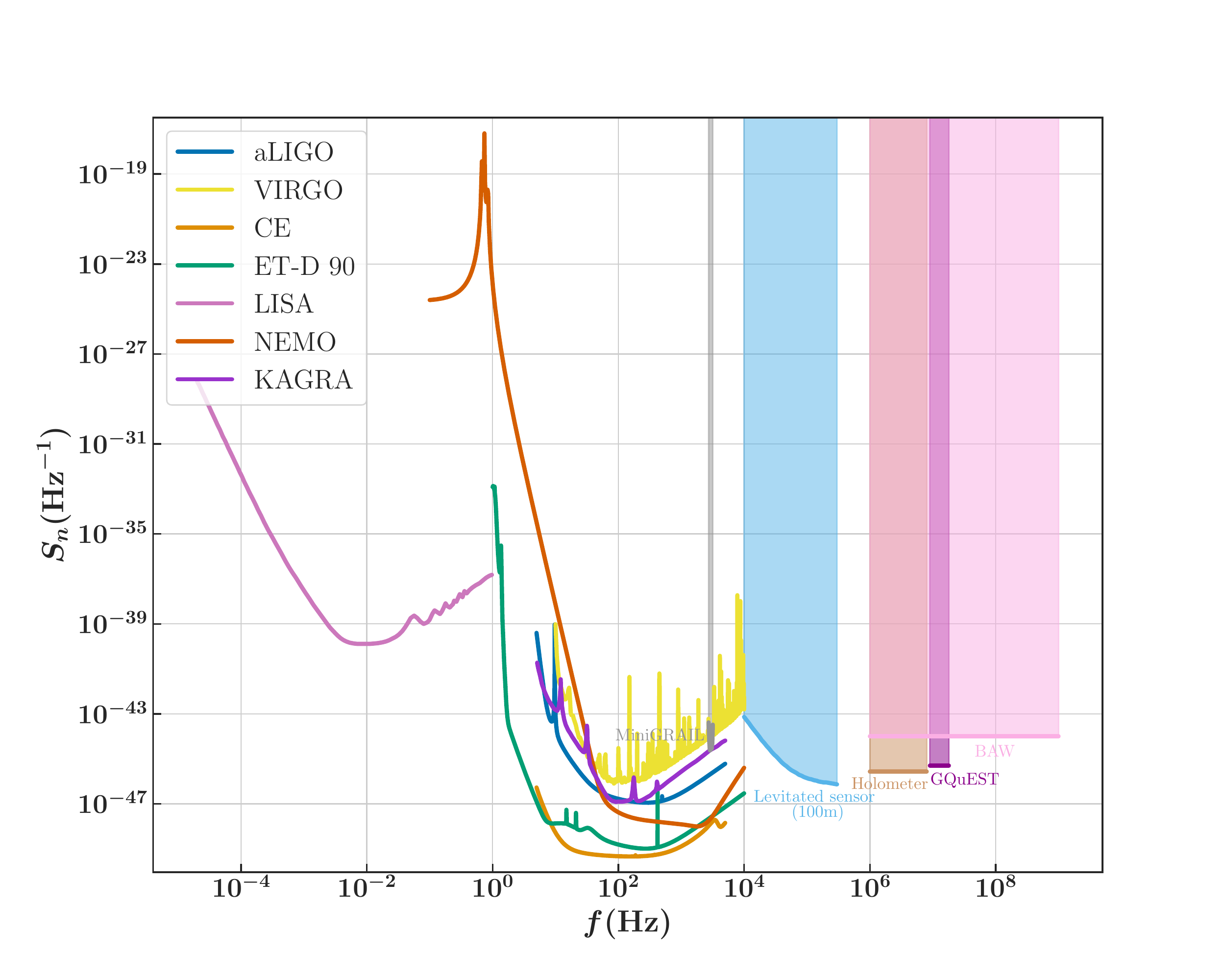} 
    \caption{Sensitivity curves of GW experiments. The projected noise spectral densities for laser interferometers are plotted in solid lines. Strain sensitivities for narrowband detectors are shown in shaded regions within the quoted bandwidths. Note that the presented frequency range for the levitated sensor corresponds to the tunable frequency range of the trapping potential, rather than the measurement bandwidth. See Table~\ref{tab:overview} for an overview of the GW experiments and Sec.~\ref{sec:sensitivity} for a description of experiments.
    }\label{fig:Sh}
\end{figure*}

In this section, we discuss various types of GW experiments that are sensitive to transiting macroscopic DM signals, with a focus on laser interferometers.  
An overview of experiments considered in this paper is given in Table~\ref{tab:overview}.
A collection of noise spectral densities for such experiments can be found in Fig.~\ref{fig:Sh}. 

\renewcommand{\arraystretch}{1.5}
\begin{center}
\begin{table}[!ht] 
    \centering
    \begin{tabular}{|c||c|c|c|c|}
        \hline
        \hline 
        Experiment    
        & Operating Frequency \footnote{For narrowband experiments, the readout bandwidth is additionally quoted in brackets.}
        & \makecell{ $S_n(f_{\rm peak})$ \\ $[\rm{Hz}]^{-1}$ }
        & \makecell{   $L $ \\ $[{\rm m}]$ }  
        & Detector Geometry \footnote{For interferometers with two arms, we quote the angle between the two arms.} \\
        \hline
        \hline
        \multicolumn{5}{|c|}{Laser interferometers} \\
        \hline
        LISA \cite{LISA}  & $[10^{-4},1]\,$Hz & $10^{-40}$ & $2.5 \times 10^9$ & $\pi/3$ \\
        LIGO (aLIGO, Voyager) \cite{aLIGO,LIGO:2020xsf} & $[10,10^4]\,$Hz & $10^{-47}$ & $4000$ & $\pi/2$ \\
        KAGRA \cite{KAGRA} & $[10,10^4]\,$Hz & $10^{-47}$ & $3000$ & $\pi/2$ \\
        Virgo \cite{Virgo} & $[10,10^4]\,$Hz & $10^{-46}$ & $3000$ & $\pi/2$\\
        ET \cite{ET-D, ET-B} & $[10,10^4]\,$Hz & $10^{-49}$ & $ 10^5$ & $\pi/3$ \\
        CE \cite{CE} & $[10,10^4]\,$Hz & $10^{-50}$ & $ 4 \times  10^4$ & $\pi/2$  \\
        NEMO \cite{NEMO} & $[1,3]\,$kHz & $10^{-48}$ & $4000$   & $\pi/2$ \\
        GQuEST \cite{McCuller:2022hum} & $1.5 \times 10^7$  ($30\,$MHz)  & $10^{-46}$ & $5$ & $\pi/2$   \\
        Holometer \cite{Holometer:2017} & $[1,8]\,$MHz &  $10^{-46}$  & $39$ & $\pi/2$ \\
        \hline
        \multicolumn{5}{|c|}{Other gravitational wave detectors} \\
        \hline
        \makecell{Optically Levitated Sensor \\ (100 m) \cite{PhysRevLett.110.071105,Aggarwal:2020umq}  } & $[10 , 300]\,$kHz ($\sim 0.1\,f$) & $[10^{-42},10^{-46}]$ & 100 & cavity\\
        Bulk Acoustic Wave \cite{Goryachev:2014yra,PhysRevLett.127.071102} & MHz - GHz ($\sim 10\,$ Hz) &  $10^{-44}$ & $\sim 10^{-3}$ & plate resonator  \\
        \makecell{ Spherical Resonant Mass \cite{Gottardi:2007zn,Aguiar:2010kn} \\ (Mini-GRAIL, Schenberg antenna) }  & $3\,$kHz ($100$ Hz) & $10^{-44}$ & 0.7 & spherical mass \\
        \makecell{ Pulsar Timing Array \\ ({\it e.g.}~SKA \cite{Rosado:2015epa}) }& $[10^{-9},10^{-7}]\,$Hz & $10^{-25}$ & $\sim10^{20}$ & timing array \\
        \hline
        \hline
    \end{tabular}
    \caption{Overview of GW experiments with potential sensitivity to macroscopic DM signals through the Newtonian potential and a fifth force. See Sec.~\ref{sec:sensitivity} for a description of experiments and Fig.~\ref{fig:Sh} for the experimental sensitivity curves.}
    \label{tab:overview}
\end{table}
\end{center}

\subsection{Laser Interferometers}

 Gravitational waves were first detected by LIGO and
Virgo \cite{LIGO2016}.
Since then, laser interferometry laboratories, both ongoing and proposed, have expanded their coverage to encompass a broader range of signal frequencies. 
At higher frequencies $(\gtrsim {\rm Hz})$, the advanced LIGO and Virgo are to be joined by Cosmic Explorer \cite{CE}, Einstein Telescope \cite{ET-B,ET-D} and proposals such as NEMO \cite{NEMO} and LIGO Voyager \cite{LIGO:2020xsf}. On the other hand, LISA \cite{LISA} is proposed to operate at lower frequencies below Hz. At the same time, experimental apparatus proposed mainly to detect quantum gravity effects, such as Holometer~\cite{Holometer:2015tus}, GQuEST~\cite{McCuller:2022hum}, and the 3D interferometer \cite{Vermeulen:2020djm}, are sensitive to signals
at high frequencies in the MHz range. While our calculation cannot be applied directly to a 3D interferometer, interesting signals from a transiting DM could in principle be induced in such a device. We leave a detailed analysis for future work.

The peak sensitivity frequency is typically set by the arm length $L$ and the quality factor $Q$ of the FP cavity (if any), $f_{\rm peak} \simeq c/(4\pi Q L)$.
Throughout the paper, for laser interferometers, we consider the sensitivity as obtained by a two-arm configuration. The angle between the two arms is given by the proposed detector geometry, as quoted in~Table~\ref{tab:overview}.
Throughout this work, we have used the published noise curves for all experiments as shown in Fig.~\ref{fig:Sh} unless otherwise specified. 

GQuEST is in principle a narrowband detector utilizing novel photon counting techniques to evade the standard quantum limit~\cite{McCuller:2022hum}, where the bandwidth and estimated sensitivity are given in~Table~\ref{tab:overview}. We note that the sensitivity of GQuEST given in Table~\ref{tab:overview} assumes an integration time of 1000 s and a single interferometer setup in which the auto-correlation is measured. Future development of a colocated setup, allowing for cross-correlation and extended operation of GQuEST, can lead to a better sensitivity than the conservative estimate given in Table~\ref{tab:overview}.

\subsection{Other Types of Gravitational Wave Detectors}

\textbf{Optically levitated sensors} with tunable frequencies have been proposed to detect GWs at high frequencies in the $\sim [10,300]\,{\rm kHz}$ range \cite{PhysRevLett.110.071105,Aggarwal:2020umq}. Such a device consists of a nanoscale or microscale sensor (sphere or disk) levitated optically and placed at an antinode of a tunable trapping laser inside an FP cavity. An interferometerlike configuration \cite{Aggarwal:2020umq} further increases the sensitivity by noise cancellation between its two arms. A one-meter prototype of the detector is under construction. The optically levitated sensor is a resonant detector, where the motion of a dielectric nanoparticle suspended at an antinode of the cavity can be detected. The operating frequency is determined by the tunable trapping frequency of the nanoparticle: $\omega_0^2 = \frac{1}{m_s} \frac{d^2 U}{d x^2} |_{x = x_0}$, where $m_s$ is the nanoparticle mass, $U$ is the optical potential, and $x_0$ is the antinode location. The dominant noise source is from the Brownian thermal motion of photon scattering from the nanoparticle,
 which is suppressed at higher frequencies and cryogenic temperatures. For such resonant sensors, we assume the test mass to be free over the interaction timescale, so the characteristic frequency of the signal is larger than the trapping frequency, $2\pi f_\tau\gtrsim\omega_0$. 

In a local Lorentz frame with the inner mirror at the origin, we treat both the levitated object and the end mirror as free objects within one measurement, initially at $x_s$ and $\ell_m$. The relevant quantity measured is the displacement of the levitated object from the antinode of the trapping laser, $\Delta x = \delta x_\text{\rm min}-\delta x_s=\delta\ell_m-\delta x_s$. A detector specialized to probe GW signals would benefit from having $x_s$ as small as possible \cite{PhysRevLett.110.071105}, so here we also work in the limit where $x_s\ll\ell_m$, treating the device as an interferometer with two mirrors separated by a distance of $\sim\ell_m$. We use the strain sensitivity quoted in Ref.~\cite{Aggarwal:2020umq} to calculate the SNR, without a careful treatment of the cavity response. The cavity response might boost the signal on the higher frequency end up to an order of magnitude (for the $100$ m stack detector). In addition to the classical Doppler acceleration, we also consider the Shapiro effect, which can displace the minimum of the optical potential and hence the sensor location. We estimate the Shapiro strain using formulas derived for laser interferometers in Sec.~\ref{subsec:shapiro}. We emphasize that, while the leading order effect in an optically levitated sensor is expected to be captured by the Doppler component of the interferometry signal, the interferometry treatment here is an approximation, where the total effects are not gauge invariant in the setup of an optically levitated sensor with a trapping potential. We leave a more detailed calculation for future work. 

We note that there are experiments with standalone levitated or trapped test masses not included in this paper \cite{Biercuk_2010,Timberlake_2019,Lewandowski_2021,PhysRevA.101.053835,Deli__2020,Monteiro2020,PhysRevLett.124.013603,Ahn_2020}. They either operate at lower frequencies comparable to laser interferometers or with lower acceleration sensitivities than the apparatus considered. 
Another recent proposal involving an array of levitated mechanical sensors and the projected sensitivity to composite DM can be found in Ref.~\cite{Carney_2020}.

\textbf{Resonant mass detectors} have their origins at the beginning of experimental GW physics, {\it i.e.}~the Weber bar experiment. In general, resonant vibration modes of the test mass as induced by an external force can be sensed through certain read-out systems. Along one direction of the test mass, considering the fundamental mode, the strain sensitivity of such detectors can be converted to acceleration sensitivity according to $\tilde{h}(f) \sim \tilde{a}(f) / (8 L f^2)$, where $L$ is the length of the resonant mass in such a direction \cite{Maggiore:2007ulw}. To estimate the DM sensitivity, we consider the classical Doppler acceleration as given by~Eq.~\eqref{eqn:acceleration_M1_fourier} projected onto one direction. We acknowledge that the exact signal induced by the DM's potential requires a careful calculation that can be done using the metric perturbation formalism. However, as the DM sensitivity for such experiments is suppressed, as shown in Figs.~\ref{fig:reach} and \ref{fig:fifth_force_reach}, we do not attempt to refine the calculation further. \textit{Spherical resonant mass} experiments, such as Mini-GRAIL \cite{Gottardi:2007zn} and Schenberg antenna \cite{Aguiar:2010kn}, transform excited mechanical modes to electrical signals. Such experiments operate at $\sim\,$kHz frequencies, and the strain sensitivities are typically less than laser interferometers operating at the same frequency. At the same time, a new type of resonant mass detectors, the 
\textit{Bulk Acoustic Wave} experiment, is designed to operate at higher frequencies in the MHz - GHz range \cite{Goryachev:2014yra,PhysRevLett.127.071102}. These experiments generally sense the acoustic waves inside a piezoelectric material
along a certain direction ({\it e.g.}, for a cylinder, along the length) through the SQUID readout. There is a broad range of operating frequencies depending on the acoustic eigenmode. The strain sensitivity is improved with a large mode quality factor and cryogenic temperatures. Note that such experiments employ higher resonance modes to achieve broadband sensitivity. To convert the strain sensitivity, we have assumed the fundamental mode.
Notice that at a similar frequency range, membrane optomechanical experiments based on optical cavities (see, {\it e.g.}, Ref.~\cite{stgelais2018sweptfrequency}) can reach a similar acceleration sensitivity of $\sim 10^{-5}\, {\rm m}/{\rm s}^2/\sqrt{\rm Hz}$. We do not make explicit projections for such experiments and refer the readers to the BAW projection as a reference.

\textbf{Long-baseline atom interferometers} are another venue for both GW and DM direct detection. There has been a growing interest in such detectors and active proposals. Long-baseline atom interferometers consisting of two spatially separated single atom interferometers are proposed to close the midband window between the low-frequency LISA ($\sim 0.01\,$Hz) and ground-based laser interferometers ($\sim 1\,$Hz). The operation frequency is limited by gravity gradient noise on the lower end and the rate of relaunching cold atoms on the higher end. With improved noise models and space-based designs, some missions can cover a lower frequency range even beyond LISA (see Ref.~\cite{2022RSPTA.38010060B} for a review). 
A long-baseline atom interferometer, such as MAGIS \cite{Graham:2017pmn,MAGIS-100:2021etm}, AEDGE \cite{AEDGE}, or AION \cite{AION}, resembles a single-arm laser interferometer that can perform differential phase measurements at any given time and benefit from noise cancellation between the two devices. Although a single atom interferometer can also serve as an accelerometer, the tidal effect is determined by the rather small wave packet separation, typically $\lesssim 1\,$m, that would generally suppress the sensitivity. Here we briefly discuss the prospects for long-baseline atom interferometers and leave detailed studies for both types of proposals for future work. A long-baseline atom interferometer precisely measures the light traveling time between the two atom interferometers distantly separated by a baseline length $L$. The two atom interferometers are run by a common laser. The laser drives the atomic transition between the ground and excited states and transfers $2 \pi \hbar /\lambda $ momentum to the atoms at each pulse, where $\lambda$ is the laser wavelength. Laser pulses applied at different times serve as ``mirrors" and the ``beamsplitter" for the interferometer. The phase of the interference fringe at each atom interferometer depends both on the laser phase and the phase accumulated by the atoms themselves. 
The pair of atom interferometers serve as both precise inertia and laser frequency reference. 
For the single-photon transition scheme, the relative interference phase between the two atom interferometers is given by $\Delta \phi = \omega_{\rm A} \, (2 L /c)$, 
where $\omega_{\rm A}$ is the atomic transition frequency, and the baseline length determines the light traveling time.
Thus, naively, the strain on the baseline length $h(t) \sim \Delta L/L$, as induced by the transiting DM interacting with the atom cloud and the traversing photon through the Newtonian potential, can be sensed by such detectors. However, the exact phase shift as induced by the transiting DM depends on the internal mechanism of the atom interferometry, as well as how the photon propagates with space-time fluctuations. For example, the Shapiro effect can be dominant in the high DM mass regime, and cannot be captured by the classical accelerometer projection based on the Doppler effect. The strain sensitivity of proposed long-baseline atom interferometers can be comparable to laser interferometers. Thus, we postpone the study of the gauge invariant phase calculation to future work.

\textbf{Pulsar timing arrays} have been studied as powerful and complementary probes to DM subhalos at small masses ($M < 10^{2} M_{\odot}$) \cite{Siegel:2007fz, Baghram:2011is, Clark:2015sha, Dror:2019twh, Ramani:2020hdo, Lee:2020wfn}. In this work, we appropriately extend the results from \cite{Lee:2020wfn} to lower masses using analytic results, assuming observations of 200 pulsars across 20 years of observational time, 2 weeks of cadence, and 50 ns of white noise in the timing data, which corresponds to an estimated scenario of the Square Kilometer Array (SKA) experiment~\cite{Keane:2014vja}.

\section{Discussion \& Conclusion}\label{sec:discussion}

In this paper, we consider the effects of transiting macroscopic dark matter on GW experiments, particularly laser interferometers. Gravitational interaction and an additional Yukawa interaction are both considered. We applied the formal gauge invariant observable for laser interferometers to the case of transiting DM.  Importantly, in addition to the Dopper effect, which is the only effect considered in existing literature, the Shapiro effect and Einstein delay may also be present for a generic interferometer design. The Shapiro effect is the change in the messenger travel time along the interferometer arm. The Einstein delay is the time dilation of the clock's proper time, which cancels between arms for a two- or multi-baseline interferometer. We show that, for most operating and proposed laser interferometers, the Shapiro effect is subdominant compared to the Doppler effect. However, we also observe that depending on the experimental parameters, the Shapiro effect may take over for higher DM masses. 

In general, GW experiments operating at higher frequencies are sensitive to macroscopic DM with lower masses. Across the landscape of experiments included in this paper, apart from PTAs in the very low-frequency range, for laser interferometers in the $10^{-4}\,{\rm Hz} - {\rm kHz}$ range and high-frequency apparatus (including Holometer and GQuEST) in the ${\rm kHz} - {\rm GHz}$ range, the projections peak at DM masses in the range of $\sim 1 - 10^{15}\,$kg. %
The signal is dominated by transiting DM with an impact parameter smaller than the interferometer baseline length, {\it i.e.},~$b < L$. This is a result of several effects, such as the peak frequency of the DM signal compared to that of the experiment, the tidal effect, and the time differential effect of the strain measurement. Typically the peak frequency is the dominant factor. However, for experiments with a large quality factor, the tidal effect may be the most relevant suppression for DM with large impact parameters. 

We have also investigated constraints from the stochastic signal produced by an ensemble of transiting DM. We found that for the Doppler effect, the constraints from stochastic signals are always weaker than the deterministic constraints. On the other hand, for the Shapiro effect, the stochastic signal could dominate over the deterministic signal in the low mass regime.

Lastly, we have left out the analysis of an important type of GW experiment, {\it i.e.}~atom interferometers, as the exact gauge invariant observable induced by transiting DM is different than that of laser interferometers. We postpone such a study for future work.  %

\begin{acknowledgments}
We thank Sebastian Baum, Mathew Bub, Yanbei Chen, Michael Fedderke, Moira Gresham, Dongjun Li, Clara Murgui, Kris Pardo, and Yiwen Zhang for helpful discussions on related topics. This work is supported by the Quantum Information Science Enabled Discovery (QuantISED) for High Energy Physics (KA2401032), the U.S. Department of Energy, Office of Science, Office of High Energy Physics, under Award No. DE-SC0011632, and by the Walter Burke Institute for Theoretical Physics. The computations presented here were conducted in the Resnick High Performance Computing Center, a facility supported by the Resnick Sustainability Institute at the California Institute of Technology.

\end{acknowledgments}

\appendix

\section{Angular Factors}
\label{app:angular_factors}

In this appendix we derive the mean values of angular factors involving dot and cross products between $\unit{v}$, $\unit{b}$ and $\unit{n}$, assuming an isotropic distribution of $\unit{v}$, and a uniform distribution of $\unit{b}$ constrained on a plane perpendicular to $\unit{v}$. We note that while the DM velocity in the lab frame has a preferred direction, the isotropic velocity distribution is a good approximation for analytic estimates, as verified by the Monte Carlo simulation.

Without loss of generality, we set $\unit{n}=(0,0,1)$ on the $z$ axis. In spherical coordinates, we write $\unit{v}=(\sin\theta\cos\phi,\sin\theta\sin\phi,\cos\theta)$. Since $\unit{b}$ is constrained to be perpendicular to $\unit{v}$, we can write $\unit{b}=\cos\varphi\unit{b}_1+\sin\varphi\unit{b}_2$, where $\unit{b}_1=(-\sin\phi,\cos\phi,0)$ and $\unit{b}_2=(\cos\theta\cos\phi,\cos\theta\sin\phi,-\sin\theta)$ are orthogonal unit vectors that are perpendicular to $\unit{v}$. Averages over an angular factor $X$ are then computed by the integral
\begin{equation}\label{eqn:angular_general}
	\langle X\rangle =\frac{1}{8\pi^2}\int_0^{\pi}\sin\theta d\theta\int_0^{2\pi}d\phi\int_0^{2\pi}d\varphi\,  X(\theta,\phi,\varphi) \, .
\end{equation}
One easily evaluates $\unit{v}\cdot\unit{n}=\cos\theta$,  $\unit{b}\cdot\unit{n}=-\sin\theta\sin\varphi$, $|\unit{v}\times\unit{n}|^2=\sin^2\theta$ and $|\unit{b}\times\unit{n}|^2=\cos^2\phi+\cos^2\theta\sin^2\phi$, giving
\begin{align}\label{eqn:angular_mean}
	\langle(\unit{v}\cdot\unit{n})^2\rangle &= \langle(\unit{b}\cdot\unit{n})^2\rangle =\frac{1}{3}\nonumber \\ 
	\langle|\unit{v}\times\unit{n}|^2\rangle &= \langle|\unit{b}\times\unit{n}|^2\rangle =\frac{2}{3}  \, .
\end{align}
Cross terms between $\unit{b}$ and $\unit{v}$ include $(\unit{b}\cdot\unit{n})(\unit{v}\cdot\unit{n})=-\sin\theta\cos\theta\sin\varphi$ and $(\unit{b}\times\unit{n})\cdot(\unit{v}\times\unit{n})=\sin\theta\cos\theta\sin\varphi$, which integrate to zero
\begin{align}\label{eqn:angular_mean_cross}
	\langle(\unit{b}\cdot\unit{n})(\unit{v}\cdot\unit{n})\rangle &=0\nonumber \\ 
	\langle(\unit{b}\times\unit{n})\cdot(\unit{v}\times\unit{n})\rangle &=0  \, .
\end{align}
Higher-order factors include $(\unit{b}\cdot\unit{n})^4=\sin^4\theta\sin^4\phi$, $(\unit{v}\cdot\unit{n})^4=\cos^4\theta$ and $(\unit{b}\cdot\unit{n})^2(\unit{v}\cdot\unit{n})^2=\sin^2\theta\cos^2\theta\sin^2\varphi$,
\begin{align}\label{eqn:angular_four}
	\langle(\unit{b}\cdot\unit{n})^4\rangle = \langle(\unit{v}\cdot\unit{n})^4\rangle &=\frac{1}{5}\nonumber \\ 
	\langle(\unit{b}\cdot\unit{n})^2(\unit{v}\cdot\unit{n})^2\rangle &=\frac{1}{15} \, .
\end{align}
For interferometers with two arms separated by $\Delta\theta$, we set $\unit{n}_1=(0,0,1)$ and $\unit{n}_2=(0,\sin\Delta\theta,\cos\Delta\theta)$, again without loss of generality. Then the angular differences between the two arms are
\begin{align}\label{eqn:angular_twoarn}
	\langle[(\unit{b}\cdot\unit{n}_1)-(\unit{b}\cdot\unit{n}_2)]^2\rangle = 	\langle[(\unit{v}\cdot\unit{n}_1)-(\unit{v}\cdot\unit{n}_2)]^2\rangle &=\frac{4}{3}\sin^2\left(\frac{\Delta\theta}{2}\right)\nonumber \\ 
	\langle[(\unit{b}\cdot\unit{n}_1)^2-(\unit{b}\cdot\unit{n}_2)^2]^2\rangle = 	\langle[(\unit{v}\cdot\unit{n}_1)^2-(\unit{v}\cdot\unit{n}_2)^2]^2\rangle &=\frac{4}{15}\sin^2\Delta\theta\nonumber \\ 
	\langle[(\unit{b}\cdot\unit{n}_1)(\unit{v}\cdot\unit{n}_1)-(\unit{b}\cdot\unit{n}_2)(\unit{v}\cdot\unit{n}_2)]^2\rangle  &=\frac{1}{5}\sin^2\Delta\theta \, .
\end{align}

\bibliography{bib}{}
\bibliographystyle{JHEP}

\end{document}